\documentclass[12pt]{article}
\usepackage[english]{babel}
\usepackage{graphicx}
\usepackage{framed}
\usepackage{color}
\usepackage[normalem]{ulem}
\usepackage{amsmath}
\usepackage{amsthm}
\usepackage{amssymb}
\usepackage{amsfonts}
\usepackage{enumerate}
\usepackage[utf8]{inputenc}
\usepackage[top=1 in,bottom=1in, left=1 in, right=1 in]{geometry}
\usepackage{hyperref}
\usepackage{natbib}
\usepackage{multirow}
\usepackage{caption}
\usepackage{subcaption}

\theoremstyle{definition}
\newtheorem{theorem}{Theorem}

\newtheorem{proposition}{Proposition}
\newtheorem{assumption}{Assumption}

\newtheorem{remark}{Remark}

\title{Tail Gini Functional under Asymptotic Independence}
\author{Zhaowen Wang\\
    Department of Statistics and Data Science, School of Management,\\ Fudan University\\
    Liujun Chen \\
    International Instutite of Finance, School of Management, \\
    University of Science and Technology of China\\and \\
    Deyuan Li \\
    Department of Statistics and Data Science, School of Management,\\ Fudan University}

\date{}
\begin{document}\maketitle
Tail Gini functional is a measure of tail risk variability for systemic risks, and has many  applications in banking, finance and insurance. Meanwhile, there is growing attention on aymptotic independent pairs in quantitative risk management. This paper addresses the estimation of the tail Gini functional under asymptotic independence. We first estimate the tail Gini functional at an intermediate level and then extrapolate it to the extreme tails. The asymptotic normalities of both the intermediate and extreme estimators are established. The simulation study shows that our estimator performs comparatively well in view of both bias and variance. The application to measure the tail variability of weekly loss of individual stocks given the occurence of extreme events in the market index in Hong Kong Stock Exchange provides meaningful results, and leads to new insights in risk management.

\section{Introduction}
\label{sec:intro}

Meauring tail risk is crucial in many different fields such as banking, finance and insurance. The most popular tail-based risk measures in the literature are Value at Risk (VaR) and Expected Shortfall (ES). See \cite{patton2019dynamic}, \cite{li2022pelve} and \cite{hoga2022monitoring} for some recent discussions on VaR and ES in quantitative risk management. It is worth noticing that both VaR and ES fail to capture the variability of tail risk. Variability has been a prominent notion since the publication of the seminal work \cite{markowitz1952portfolio}. Variance, standard deviation and Gini mean difference are classical measures of variability. Among them, Gini mean difference, first introduced by \cite{gini1912variabilita}, has been hugely influential in numerous research areas \citep[e.g.][]{yitzhaki2013gini}.  Gini mean difference can be written in the form of a covariance: 
$$
\operatorname{Gini}(X)=4 \operatorname{Cov}\left(X, F_1(X)\right),
$$
where $F_1$ is the cumulative distribution function of random variable $X$. 

To incorporate variability in tail risk analysis, \cite{furman2017gini} extends the (classic) Gini mean difference to tail Gini functional. The tail Gini functional for $X$ is defined by $$
\operatorname{TG}_p(X)=\frac{4}{p} \operatorname{Cov}\left(X, F_1(X) \mid F_1(X)>1-p\right),
$$ where $p>0$ is a sufficient small value. Obviously, the tail Gini functional $
\operatorname{TG}_p(X)$ is designed to measures the variability of $X$ on the upper tail region and quantifies the risk of $X$ solely. In finance, $X$ may be the loss of an individual asset and tail Gini functional for $X$ indicates the risk measure of tail variability of that asset. 

However, in practice, regulators are concerned not only with measuring risks for individual asset or  entity, but also with measuring individual risks given the impact from systemic variables. A series of studies have delved into modeling and measuring systemic risks, including \cite{cai2015estimation}, \cite{tobias2016covar}, \cite{acharya2017measuring}, and so on. To extend tail Gini functional for systemic risk analysis, we consider tail Gini functional for bivariate random vector $(X, Y)$, introduced by \cite{hou2021extreme} as $$
\operatorname{TG}_p(X ; Y)=\frac{4}{p} \operatorname{Cov}\left(X, F_2(Y) \mid F_2(Y)>1-p\right),
$$
where $F_2$ is the marginal cumulative distribution function of $Y$. Here, $Y$ could be a systemic variable indicating the loss of a financial system. By conditioning on $F_2(Y)>1-p$, we are focusing on the the tail variability of $X$ under the tail scenarios of the systemic variable $Y$. In this way, $\operatorname{TG}_p(X; Y)$ is a systemic tail variability measure incorporating both the marginal risk severity of $X$ and the tail structure of $(X, Y)$. 

Under the assumption that $X$ is in the Fr\'echet domain of attraction with the extreme value index $\gamma_1\in (0,1)$, \cite{hou2021extreme} obtains the asymptotic limit of $\operatorname{TG}_p(X; Y)$, that is $$\lim _{p \rightarrow 0} \frac{\operatorname{TG}_p(X; Y)}{Q_1(1-p)}=\theta_0\in[0, \infty),
$$where $Q_1$ is the quantile function of $X$. The condition $0<\gamma_1<1$ guarantees that $\mathbf{E}[X]$ exsits. Based on the above limit,  \cite{hou2021extreme} proposes an estimator for $\operatorname{TG}_p(X; Y)$ with $\theta_0>0$. Note that the corresponding $\theta_0>0$ holds only if $X$ and $Y$ are asymptotically dependent. If $X$ and $Y$ are asymptotically independent, we have $\theta_0=0$. In this paper, we will study the estimation of $\operatorname{TG}_p(X; Y)$ under asymptotic independence. We refer to \cite{ledford1996statistics} for the concepts of asymptotic independence and asymptotic dependence.

Although most research articles in bivariate extreme value framework deal with asymptotic dependence, there is increasing evidence that weaker dependence actually exists in bivariate tail region in many applications, for example, significant wave height \citep{wadsworth2012dependence}, spatial precipitation \citep{le2018dependence} and daily stock prices \citep{lehtomaa2020asymptotic}. Asymptotic independence is therefore the more appropriate model for such applications. In the field of quantitative risk management, there is also growing attention on risk measures for asymptotically independent pairs \citep[see][etc]{kulik2015heavy,das2018risk,cai2020estimation,sun2022extreme}. 

\cite{ledford1996statistics} proposes the coefficient of tail dependence, named $\eta$, to measure the severity  asymptotic independence. Assume that there exists an $\eta \in(0,1]$ such that the following limit exists and is positive for all $(x, y) \in(0, \infty)^2$:
\begin{equation}
    \lim _{p \rightarrow 0} p^{-\frac{1}{\eta}}  \mathbf{P}\left(1-F_1(X)<px, 1-F_2(Y)<py\right)=: \tau(x, y)>0.\label{1}
\end{equation}
The coefficient of tail dependence $\eta$ describes the strength of extremal dependence in the bivariate tail. If $\eta=1$, we say that $X$ and $Y$ are asymptotically dependent. If $0<\eta<1$, we say that $X$ and $Y$ are asymptotically independent. Moreover, if $1/2<\eta<1$, $X$ and $Y$ are called asymptotically independent but positively associated; if $0<\eta<1/2$, $X$ and $Y$ are called asymptotically independent but negatively associated. When $X$ and $Y$ are independent, then $\eta=1/2.$ For more details on the interpretation of $\eta$ see \cite{ledford1996statistics}.

It is the goal of this paper to estimate  $\operatorname{TG}_p(X; Y)$ under asymptotic independence but positive association ($1/2<\eta<1$). To our best knowledge, there is no literature addressing the estimation problem for  tail-based measure of variability for asymptotically independent structures. Our work is also of great significance since we considers not only positive loss variables but also real loss variables in the context of asymptotic independence. For the case of asymptotic independence but negative association ($0<\eta<1/2$), it is much more technically challenging and we leave it for future research. Meanwhile, $\eta>1/2$ is more common in real cases, see Table \ref{tb3} in Section \ref{sec4}.

The rest of the paper is organized as follows. Section \ref{sec2} studies the 
asymptotic normality of the proposed estimator for $\operatorname{TG}_p(X; Y)$. The performance of our proposed estimator is illustrated by a simulation study in Section \ref{sec3}, and a real application to Hong Kong Stock Exchange is given in Section \ref{sec4}. 
The proofs of the main theorems are provided in Section \ref{sec6}. Additional proofs are given in the supplementary material.

Throughout the paper, the notation $a_n\sim b_n$ means that $a_n/b_n\to 1$ as $n\to\infty$.  A Lebesgue measurable function $f: \mathbb{R}^{+} \rightarrow \mathbb{R}^{+}$ is called regularly varying (at infinity) with index $\alpha \in \mathbb{R}$, if
$
\lim _{t \rightarrow \infty} f(t x)/f(t)=x^\alpha,$ for $ x>0 .
$

\section{Main results}
\label{sec2}
Let $(X, Y)$ be a pair of random loss variables. 
We  propose  estimators of the tail Gini functional $\operatorname{TG}_p(X; Y)$ by a two-step approach. More specifically, we first estimate $\operatorname{TG}_p(X; Y)$ at intermediate level $p = k/n$, where $k=k(n)\to\infty$ and $k/n\to 0$ as $n\to\infty$, then extrapolate these estimators to extreme level $p =p(n)\to 0$ and $np = O(1)$ as $n\to\infty$.

Below we present our assumptions on the tail distribution of $X$ and the tail dependence between $X$ and $Y$. 
 Define $$\tau_p(x, y):=p^{-1/\eta} \mathbf{P}\left(1-F_1(X)<px, 1-F_2(Y)<py\right).$$ 
 
\begin{assumption}\label{asm1}
    There exist $\gamma_1>0$ and a regularly varying function $A_1$ with index $ \rho_1<\frac{1}{2}-\frac{1}{2 \eta}$ such that 
    $$
    \sup _{x>1}\left|x^{-\gamma_1} \frac{Q_1(1-1/(t x))}{Q_1(1-1/t)}-1\right|=O\left\{A_1(t)\right\}, \quad\text{as }  t\to\infty.$$
\end{assumption}\begin{assumption}\label{asm2}There exists $\delta>0$ such that for $y\in [0,1]$,
$$
\left|\int_0^1 \tau(x, y) d x^{-(2+\delta) \gamma_1}\right|<\infty, \quad \text { and } \quad\left|\int_1^{\infty} \tau(x, y)^2d x^{-\gamma_1}\right|<\infty .
$$
\end{assumption} 
\begin{assumption}\label{asm3}
    There exist $\beta_1>(2+\delta) \gamma_1$ and $\xi>0$ such that as $p\to 0$,
$$
\sup _{0<x\leq 1, 0<y \leq 1}\left|\tau_p(x, y)-\tau(x, y)\right| x^{-\beta_1}=O\left(p^{\xi}\right) .
$$
\end{assumption}
\begin{assumption}\label{asm4}
    There exists $0<\beta_2<\min \left\{\xi /\left(1-\gamma_1\right), 
    \gamma_1\right\}$ such that as $p\to 0$,
$$
\sup _{1<x<\infty, 0< y \leq 1}\left|\tau_p(x, y)-\tau(x, y)\right| x^{-\beta_2}=O\left(p^{\xi}\right),
$$
where $\xi$ is the same as in Assumption \ref{asm3}.
\end{assumption}
Assumption \ref{asm1} is a second order condition for 
the distribution of  $X$, which is commonly assumed in extreme value theory. We refer readers to Section 2 in \cite{de2007extreme} for the explanation of this assumption.
Assumption \ref{asm2} is a technical condition which imposes some integrality condition on the function $\tau$. Assumption \ref{asm2} and the monotonicity of $\tau(x, y)$ imply that for $\rho \in\{1,2,2+\delta\}$, 
$$
\sup _{y \in[0,1]}\left|\int_0^{\infty}\tau(x, y) d x^{-\rho \gamma_1}\right|<\infty .
$$We will deal with such integral throughout the proofs. One important property of $\tau$ is that $\tau$ is a homogeneous function of order $1/\eta$, i.e., $\tau(ax,ay) =a^{1/\eta} \tau(x,y)$ for $a>0$. We refer to \cite{lalancette2021rank} for the estimation of $\tau$. 
Assumptions \ref{asm3} and \ref{asm4} are second order strengthening of relation (\ref{1}).


\subsection{Positive loss}\label{sec2.1}
In this subsection we assume the random loss $X$ is positive.
To estimate $\operatorname{TG}_{p}(X;Y)$,
we assume that $(X_1,Y_1,),\dots,(X_n,Y_n)$ are independent copies of $(X, Y)$. A natural nonparametric estimator of $\operatorname{TG}_{k/n}(X;Y)$ at intermediate level $p=k/n\to0$ is
\begin{equation}
    \hat{\theta}_{k/n} = \frac{4n}{k^2(k-1)} \sum_{i<j} (X_i-X_j)(F_{n2}(Y_i)-F_{n2}(Y_j)) I(Y_i,Y_j>Y_{n-k,n})\label{2},
\end{equation}
where $Y_{1, n} \leq Y_{2, n}  \leq\cdots\leq  Y_{n, n}$ are the order statistics of $\{Y_1,Y_2,\cdots, Y_n\}$, and $F_{n 2}(y)=\frac{1}{n+1} \sum_i I\left(Y_i \leq y\right)$ is the empirical distribution function of $F_2$.
Moreover, we choose the intermediate sequence $k$ as follows.

\begin{assumption}\label{asm5}
    As $n \rightarrow \infty,$ $k=k(n)\to\infty$, $k/n\to 0$, $k/n^{1-\eta}\to\infty, k=O\left(n^a\right)$, where $a$ satisfies 
$$
1-\eta<a<\min \left(1-\frac{\eta}{1+\eta \gamma_1 },1+\frac{\eta}{1-2 \eta-2 \eta \gamma_1}, 1+
\frac{1}{-\frac{1}{\eta}-2 \xi+2 \beta_2\left(1-\gamma_1\right)}, 1+\frac{1}{2 \rho_1-1}
\right).
$$
\end{assumption}
Assumption \ref{asm5} imposes both lower and upper bounds for the choice of $k$. The upper bound of $k$ is a typical constraint in extreme value theory literature to control the bias of the estimators, for example see \cite{cai2020estimation}. The lower bound is used to guarantee the convergence rate $\sqrt{k}\left(\frac{n}{k}\right)^{-\frac{1}{2\eta}+\frac{1}{2}}$ goes to infinity in Proposition \ref{prop2} below.

Let $W(\cdot)$ be a mean zero Gaussian process on $[0, 1]$ with covariance structure
$$\mathbf{E}\left[W\left(y_1\right) W\left(y_2\right)\right]=-\int_0^{\infty}\tau\left(x, y_1 \wedge y_2\right) d x^{-2 \gamma_1}, \quad y_1, y_2 \in[0, 1].$$The following proposition shows the asymptotic nomality of the estimator $\hat{\theta}_{k/n}$ for $\operatorname{TG}_{k/n}(X;Y)$.
\begin{proposition}\label{prop2}Let $\{\left(X_i, Y_i\right)\}_{i=1}^n$ be independent copies of $(X, Y)$. Under the condition that $X>0$ and Assumptions \ref{asm1}-\ref{asm5},  it follows that
$$\sqrt{k}\left(\frac{n}{k}\right)^{-\frac{1}{2\eta}+\frac{1}{2}}\left(\frac{\hat{\theta}_{k/n}}{\operatorname{TG}_{k/n}(X;Y)}-1\right) \stackrel{d}{\rightarrow}\Phi:=-\frac{4}{\phi_0}\left(\int_0^1W(y)dy+\frac{1}{2}W(1)\right),$$where 
$$
    \phi_0=\frac{2(1+\gamma_1-1/\eta)}{1-\gamma_1+1/\eta}\int_0^{\infty} \tau\left(x^{-\frac{1}{\gamma_1}}, 1\right) d x.
$$
\end{proposition} 

Note that if $\eta=1$, i.e. asymptotic dependence, the convergence rate $\sqrt{k}\left(\frac{n}{k}\right)^{-\frac{1}{2\eta}+\frac{1}{2}}$ becomes $\sqrt{k}$. When $1/2<\eta<1$, i.e. asymptotic independence but positive association, the convergence rate is lower than $\sqrt{k}$.

Proposition \ref{prop2} states equivalently that
\begin{equation}\label{Eq:log}
\sqrt{k} \left(\frac{n}{k}\right)^{-\frac{1}{2\eta}+\frac{1}{2}}\log \left(\frac{\hat{\theta}_{k/n}}{\mathrm{TG}_{k / n}(X ; Y)}\right) \stackrel{d}{\rightarrow} \Phi .
\end{equation}
The log-ratio of the estimator to the true risk measure has a centered normal limit. In the simulation below, we compare the sample quantiles of log-ratios with the normal quantiles to demonstrate its asymptotic property.

Now we consider the estimation of $\operatorname{TG}_p(X ; Y)$ at extreme level $p\to0$ such that $np=O(1)$. 
\cite{sun2022extreme} shows that, for $0<\eta\leq 1$, as $n\to\infty$, 
\begin{equation}\label{prop1}
    \lim _{p \to0} \frac{\operatorname{TG}_p(X; Y)}{p^{\frac{1}{\eta}-1} Q_1(1-p)}=\phi_0 .
\end{equation}

By  \eqref{prop1}, we have that, as $n\to\infty$,
$$
\operatorname{TG}_{p}(X;Y)\sim \frac{Q_1(1-p)}{Q_1(1-k/n)} \left(\frac{k}{np}\right)^{1-1/\eta} \theta_{k/n} \sim \left(\frac{k}{np}\right)^{1-1/\eta+\gamma_1}\operatorname{TG}_{k/n}(X;Y).
$$
Thus, we  estimate $\operatorname{TG}_{p}(X;Y)$ by 
\begin{equation}
    \hat{\theta}_p = \left(\frac{k}{np}\right)^{1-1/\hat{\eta}+\hat{\gamma}_1}\hat{\theta}_{k/n}, \label{3}
\end{equation}
where $\hat{\eta}$ and $\hat{\gamma}_1$ are some suitable estimators for $\eta$ and $\gamma_1$, respectively. 

Let $k_1$ and $k_2$ be two intermediate sequences for the estimators $\hat{\eta}$ and $\hat{\gamma}_1$, respectively, i.e. $k_1=k_1(n)\to\infty$, $k_1/n\to 0$, $k_2=k_2(n)\to\infty$, $k_2/n\to 0$, as $n\to\infty$. We estimate $\gamma_1$ by the Hill estimator
$$\hat{\gamma}_1=\frac{1}{k_1}\sum_{i=1}^{k_1} \log X_{n-i+1,n}-\log X_{n-k_1,n},$$
where $X_{1,n}\leq X_{2,n}\leq\cdots\leq X_{n,n}$ are the order statistics of $\{X_{1}, X_{2},\cdots,X_{n}\}$, and estimate $\eta$ by the estimator proposed by \cite{draisma2004bivariate}
$$
\hat{\eta}=\frac{1}{k_2}\sum_{i=1}^{k_2} \log T_{n-i+1,n} -\log T_{n-k_2,n},
$$
where $T_{1,n}\leq T_{2,n}\leq\cdots\leq T_{n,n}$ are the order statistics of the non-independent but identically distributed
sequence $\{T_{1}, T_{2},\cdots,T_{n}\}$ with
$$
T_i  :=\frac{1}{(1-F_{n1}(X_i))\lor (1-F_{n2}(Y_i))}, \quad i=1,2,\dots,n, 
$$and $F_{n1}(x)=\frac{1}{n+1} \sum_i I\left(X_i \leq x\right)$ is the empirical distribution function of $F_1$.

Note that the intermediate sequences $k_1$ and $k_2$ might be different from $k$. In the rest of this paper, we choose suitable $k_1$ and $k_2$ such that
\begin{equation}\label{Eq:gamma_eta}
        \sqrt{k}\left(\hat{\gamma}_1-\gamma_1\right)=O_{\mathbf{P}}(1) , \quad \sqrt{k}\left(\hat{\eta}-\eta\right)=O_{\mathbf{P}}(1).
    \end{equation}    
Condition \eqref{Eq:gamma_eta} can be achieved by choosing $k_1$ and $k_2$ at the same order as $k$, combining with some mild conditions. We refer to Theorem 3.2.5 and Theorem 7.6.1 in \cite{de2007extreme} for the asymptotic behaviours of $\hat{\gamma}_1$ and $\hat{\eta}$, respectively.

To derive the the asymptotic normality of $\hat{\theta}_p$ at extreme level $p$, we require the following condition on the speed of $p\to0$.  \begin{assumption}\label{asm6} $\lim _{n \rightarrow \infty}(n / k)^{1 / 2-1 /(2\eta)} \log \left(d_n\right)=0$, where $d_n=k /(n p) \geq 1$. \end{assumption} 

\begin{theorem}\label{thm1}Assume the same assumptions as in Proposition \ref{prop2}. Suppose \eqref{Eq:gamma_eta} and Assumption \ref{asm6} hold. Then, as $n\to\infty$,
$$
\sqrt{k}\left(\frac{n}{k}\right)^{-\frac{1}{2 \eta}+\frac{1}{2}}\left(\frac{\hat{\theta}_p}{\operatorname{TG}_p(X; Y)}-1\right) \stackrel{d}{\rightarrow} \Phi,
$$where $\Phi$ is the same as in Proposition \ref{prop2}.
\end{theorem} 
\subsection{General Loss}In this subsection, we extend the results in Section \ref{sec2.1} to the case when the random loss $X$ is real. Denote $X^{+}=\max (X, 0)$ and $X^{-}=\min (X, 0)$, so $X=X^{+}+X^{-}$ and $\mathrm{TG}_p(X ; Y)=\mathrm{TG}_p(X^{+} ; Y)+\mathrm{TG}_p(X^{-} ; Y).$

For a real random loss $X$, we need to modify the estimator at intermediate level $p=k/n$ in (\ref{3}) as\begin{equation}\hat{\theta}_{k/n}:=\frac{4 n}{k^2(k-1)} \sum_{i<j}\left(X_i-X_j\right)\left(F_{n 2}\left(Y_i\right)-F_{n 2}\left(Y_j\right)\right)I\left(X_i, X_j>0, Y_i, Y_j>Y_{n-k, n}\right).\label{4}\end{equation}Notice this is indeed the same estimator in the case of positive random loss $X$. Therefore, we do not use another symbol to represent the estimator for the sake of simplicity. 

Under asymptotic dependence, the results for general loss could be easily derived under some mild conditions on the negative part of the general loss, see \cite{hou2021extreme}. But in the case of asymptotic independece, there is greater probability for $X$ to take negative values given large values of $Y$. This is totally different from the case of asymptotic dependence. It means that the tail variability of a general loss $X$ may not be neglectible as in the case of asymptotic dependence when the level $p$ goes to zero. 
In order to render $\mathrm{TG}_p(X^{-} ; Y)$ ignorable, we need additional conditions.\begin{assumption}\label{asm7} There exists $\zeta>1$ such that $\mathbf{E}\left|X^{-}\right|^{\zeta}<\infty$.\end{assumption} \begin{assumption}\label{asm8} $1-\frac{1}{\eta}>\xi-\beta_2$ and $\sqrt{k}(\frac{n}{k})^{-\frac{1}{2\eta}+\frac{1}{2}}p^{(1-\frac{1}{\zeta})(\frac{1}{\eta}-\beta_2+\xi)-1-\frac{1}{\eta}+\gamma_1}\to0$.\end{assumption} 

\begin{remark} Assumption \ref{asm7} imposes the condition on the left tail of $X$. Assumption \ref{asm8} is a technical condition to be used in the proof of Theorem \ref{thm2} below. 
\end{remark}Now we can apply extrapolation techniques to define $\hat{\theta}_p$ at extreme level $p$ based on the same representation (\ref{3}) with using $\hat{\theta}_{k/n}$ in (\ref{4}) instead. The asymptotic normality of $\hat{\theta}_p$ is also guaranteed.\begin{theorem}\label{thm2}Let $\{\left(X_i, Y_i\right)\}_{i=1}^n$ be independent copies of $(X, Y)$. Under the condition that $X$ is real, Assumptions \ref{asm1}-\ref{asm8} and Condition \eqref{Eq:gamma_eta}, it follows that$$\sqrt{k}\left(\frac{n}{k}\right)^{-\frac{1}{2 \eta}+\frac{1}{2}}\left(\frac{\hat{\theta}_p}{\operatorname{TG}_p(X; Y)}-1\right) \stackrel{d}{\rightarrow} \Phi,$$where $\Phi$ is the same as in Theorem \ref{thm1}.\end{theorem}
\section{Simulation}\label{sec3}In this section, we study the finite sample performance of our estimator $\hat{\theta}_p$ by simulation. We simulate the data from the following two models in \cite{cai2020estimation}. Let $a_1, a_2\in(0,1)$.

Model 1. Let $Z_1, Z_2$, and $Z_3$ be independent Pareto random variables with parameters $a_1, a_2$, and $a_1$, respectively. Here, a Pareto distribution with parameter $a>0$ means that the probability density function is $f(x)=a^{-1}x^{-1 / a-1}$ for $x>1$. Define
$$
(X, Y)=B\left(Z_1, Z_3\right)+(1-B)\left(Z_2, Z_2\right),
$$
where $B$ is a $\text{Bernoulli}(1/2)$ random variable independent of $Z_i$'s. For this model, we have $\gamma_1=a_1$, $\rho_1=1-a_1 / a_2, \eta=a_2/a_1$, and $\tau(x, y)=2^{a_1 / a_2-1}(x \wedge y)^{a_1 / a_2}$. We consider four settings of $(a_1, a_2)$, see Table \ref{tb1}.

Model 2. Define $$(X, Y)=\left((1-\Phi(\widetilde{X}))^{-a_1}, \widetilde{Y}\right),$$ where $\widetilde{X}$ and $\widetilde{Y}$ are two standard normal random variables with correlation $a_2$, and $\Phi$ is the distribution function of $\widetilde{X}$. Thus, $X$ follows from a Pareto distribution with parameter $a_1$, and $(X, Y)$ has a Gaussian copula. For this model, $\gamma_1=a_1, \rho_1=0$, $\eta=\left(1+a_2\right) / 2$, and $\tau(x, y)=(x y)^{1 /\left(1+a_2\right)}$. Obviously, $\int_0^{\infty} \tau\left(x^{-\frac{1}{\gamma_1}}, 1\right) d x=\infty$. Thus Model 2 does not satisfy Assumption \ref{asm2} and hence Theorem \ref{thm1} does not hold.

For comparison between the estimators and true values, we evaluate the true value $\mathrm{TG}_p(X ; Y)$ by using the true density functions and drawing 200 replications with sample size 1,000,000. The true values are then approximated by the corresponding median of overall 200 replications. Table \ref{tb1} shows the parameters for the distributions and the approximated true values of the tail Gini functional.

\begin{table}[]\centering
  \caption{Parameters of five models and the approximated true values of the tail Gini functional.}
  \begin{tabular}{lllllll}
  \hline
             & $(a_1, a_2)$ & $\gamma_1$ & $\eta$ & $-1/\eta+1+\gamma_1$ & $p=0.01$ & $p=0.001$ \\ \hline
  Model 1(a) & (0.35, 0.3)  & 0.35       & 6/7    & 0.183                & 0.5835  & 0.8965    \\
  Model 1(b) & (0.4, 0.35)  & 0.4        & 0.875  & 0.251                & 1.0923   & 1.9283    \\
  Model 1(c) & (0.6,0.5)    & 0.6        & 5/6    & 0.1                  & 4.2418   & 10.9131   \\
  Model 1(d) & (0.5, 0.4)   & 0.5        & 0.8    & 0.3                  & 1.3009   & 2.1104    \\ 
  Model 2 & (0.6, 0.9)   & 0.6        & 0.95   & 0.547                & 24.6808   & 84.0422    \\
  \hline
  \end{tabular}
  \label{tb1}
\end{table}
Next, we draw $m=2000$ replications from each model with sample size $n=5000$. For each replication, we compute the proposed nonparametric estimator $\hat{\theta}_p$ with $p=0.01$ and $0.001$. The proper choice of $(k, k_1, k_2)$, that is, the number of tail observations used in the estimation of $\mathrm{TG}_{k/n}(X ; Y)$, $\gamma_1$, and $\eta$, respectively, is always a delicate problem in the extreme value theory. To investigate how sensitive our result is with respect to the choice of $(k, k_1, k_2)$ and to see the range of suitable $(k, k_1, k_2)$, we compute the scaled mean squared errors $
(\operatorname{sMSE})$:
$$
\operatorname{sMSE}\left(k, k_1, k_2\right)=\frac{1}{m} \sum_{i=1}^m\left(\frac{\hat{\theta}_{p, i}\left(k, k_1, k_2\right)}{\operatorname{TG}_p(X; Y)}-1\right)^2 .
$$

\begin{figure}
    \centering
    \begin{subfigure}[b]{0.28\textwidth}
        \centering
        \includegraphics[width=\linewidth]{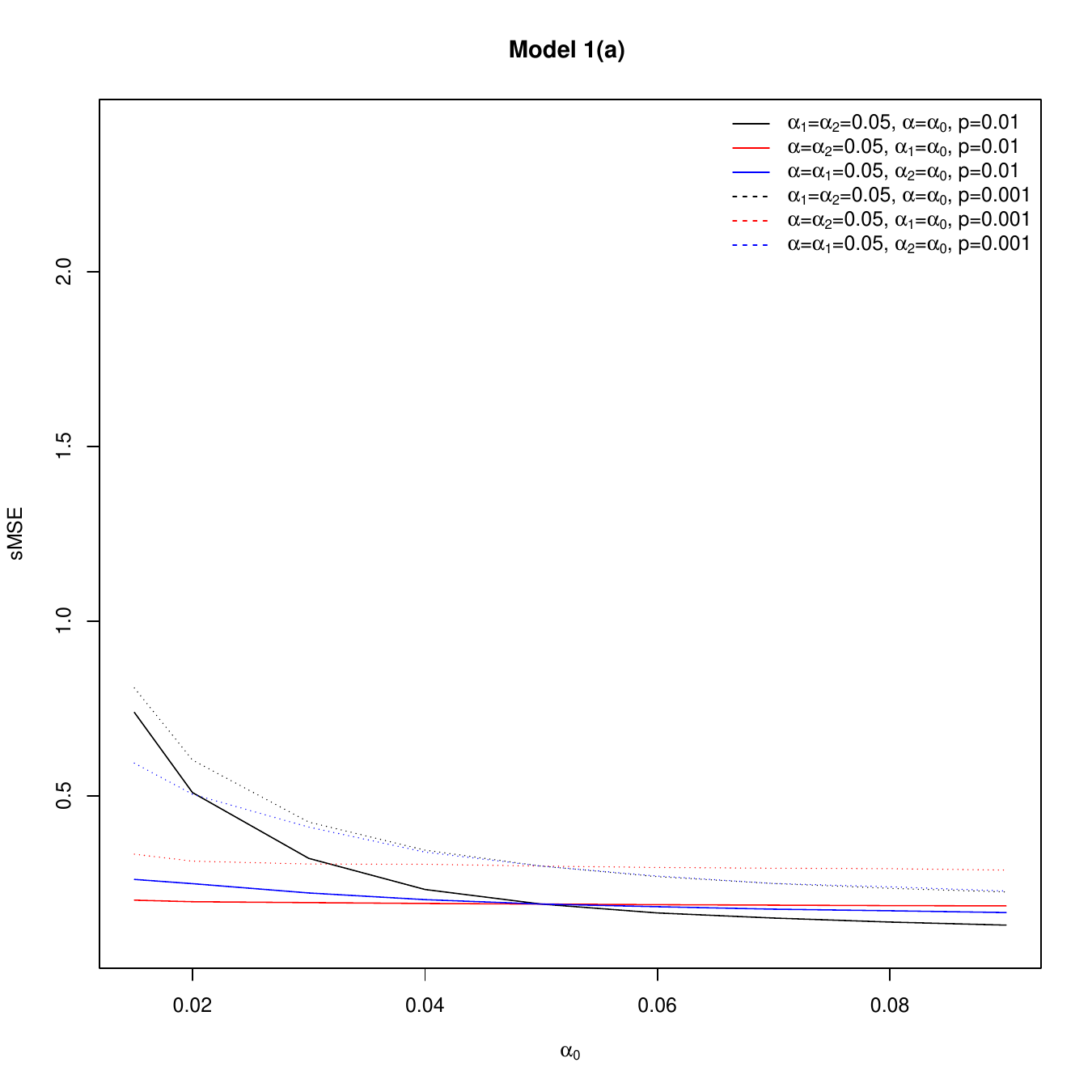}
    \end{subfigure}
    \hfill
    \begin{subfigure}[b]{0.28\textwidth}
        \centering
        \includegraphics[width=\linewidth]{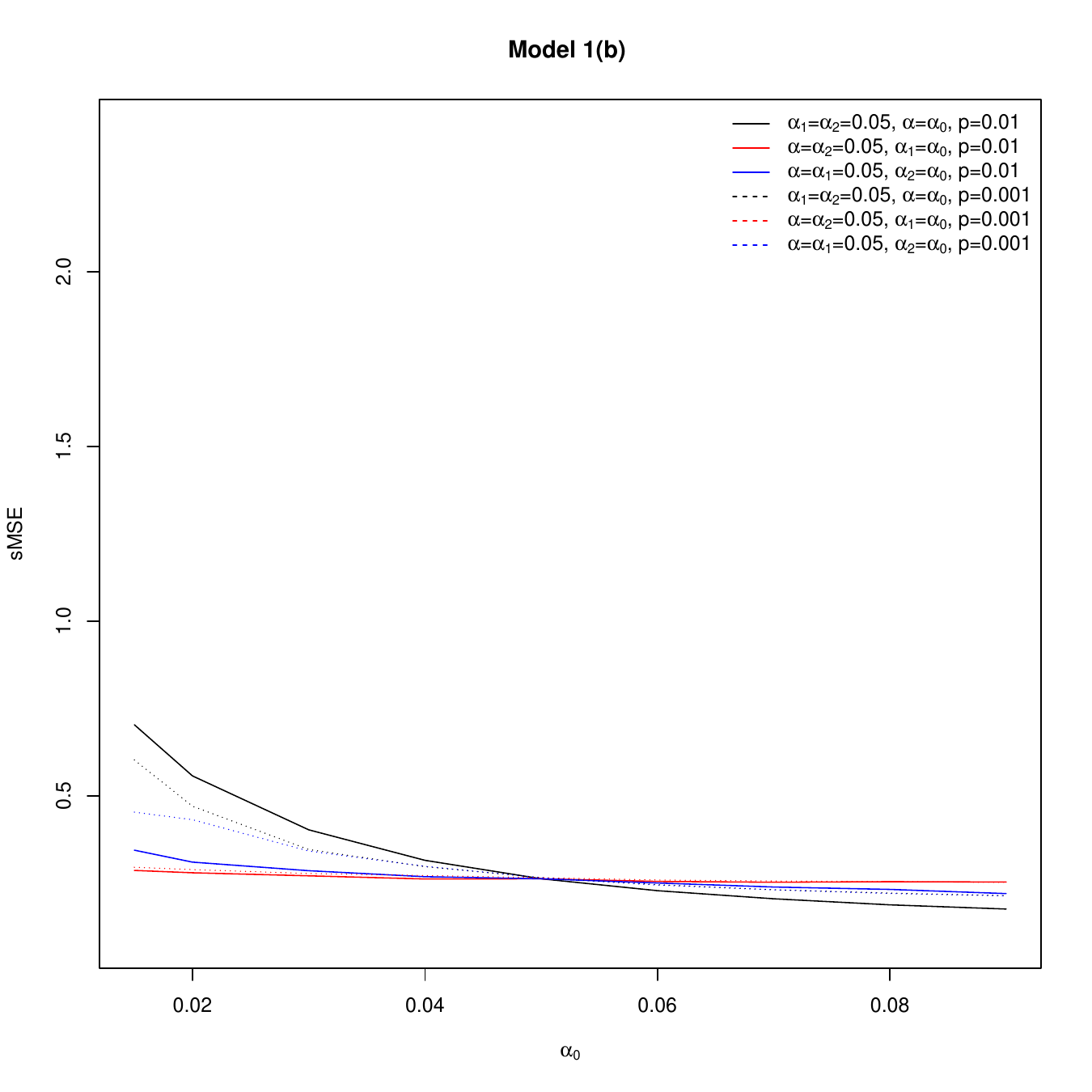}
    \end{subfigure}
    \hfill
    \begin{subfigure}[b]{0.28\textwidth}
        \centering
        \includegraphics[width=\linewidth]{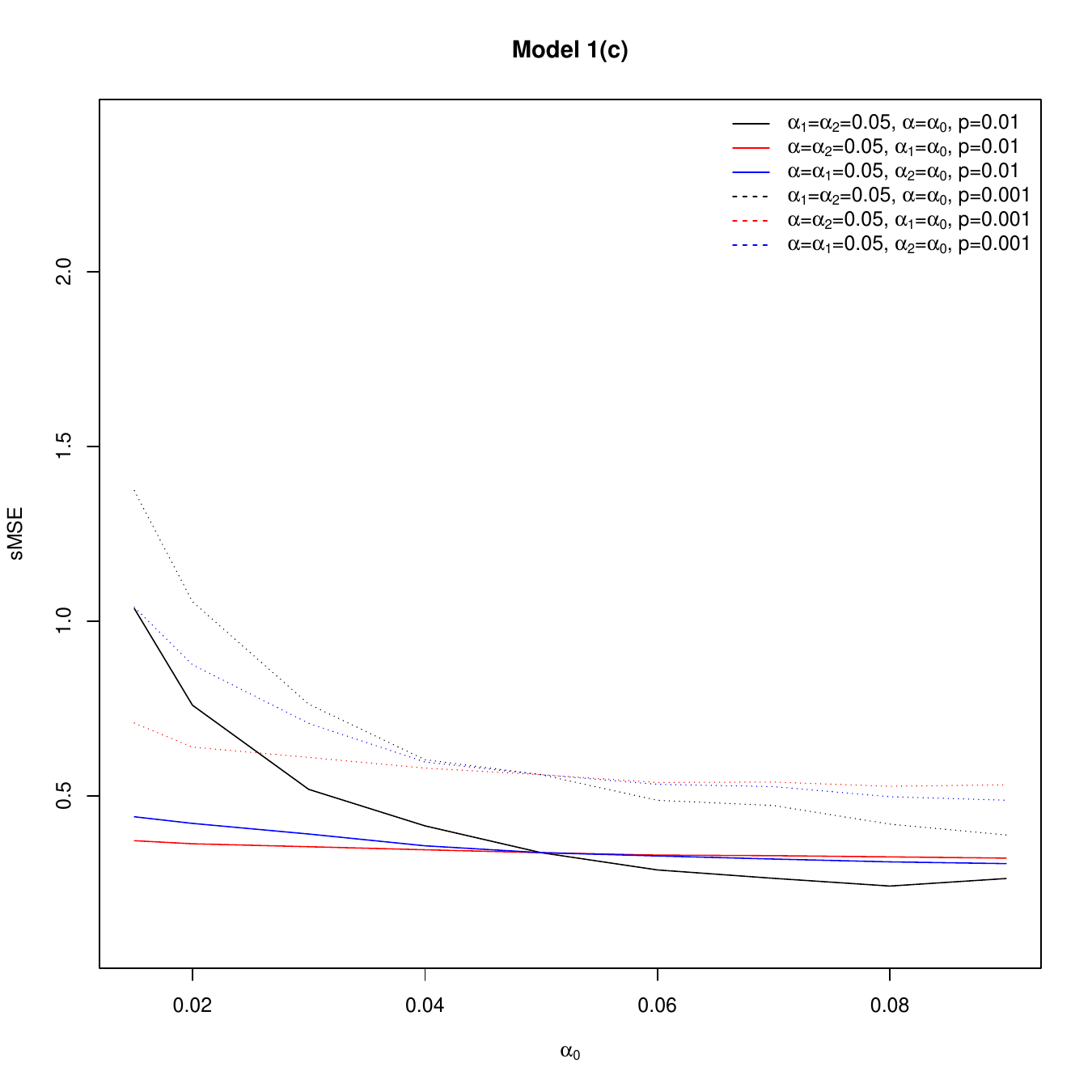}
    \end{subfigure}\hfill
    \begin{subfigure}[b]{0.28\textwidth}
        \centering
        \includegraphics[width=\linewidth]{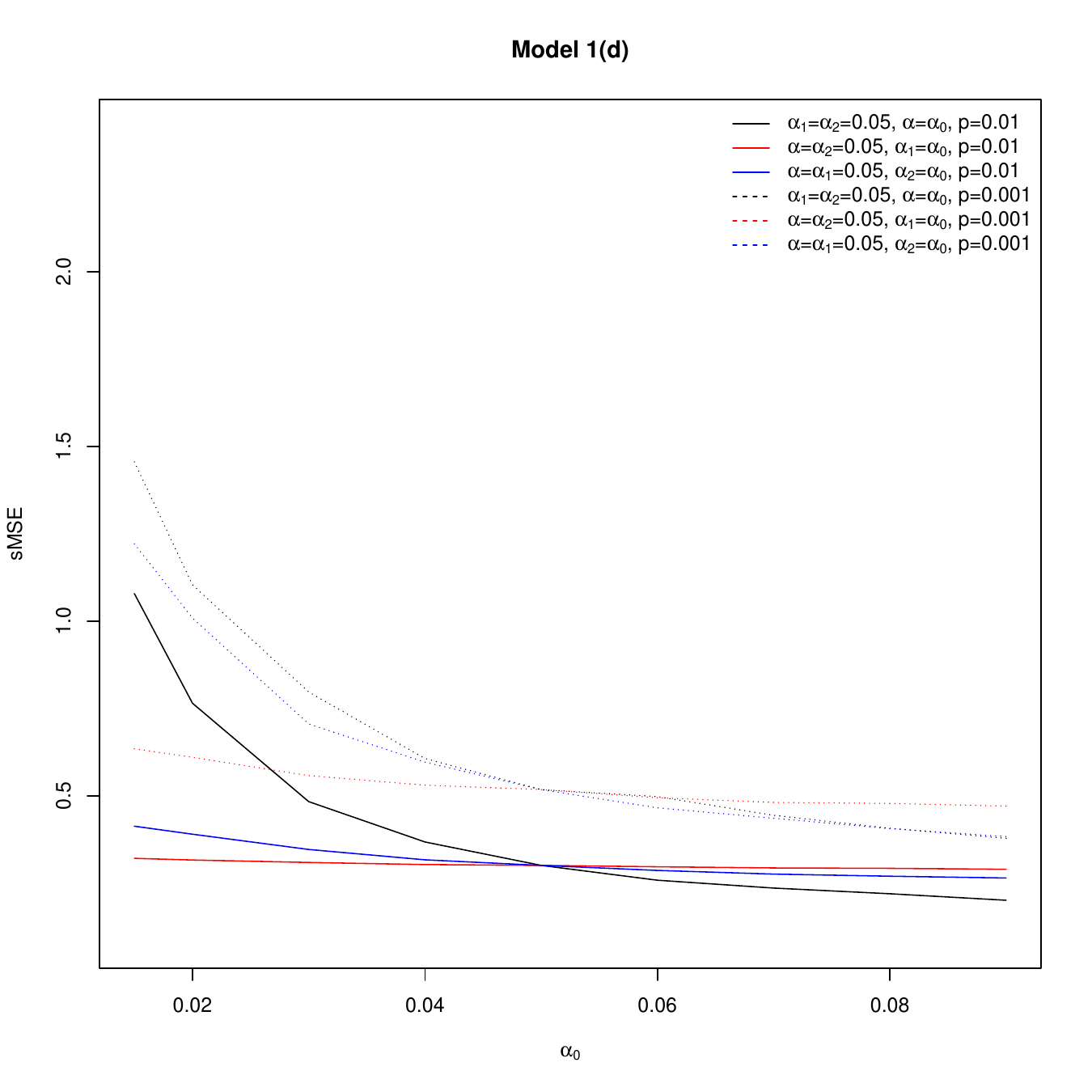}
    \end{subfigure}
    \hfill
    \begin{subfigure}[b]{0.28\textwidth}
        \centering
        \includegraphics[width=\linewidth]{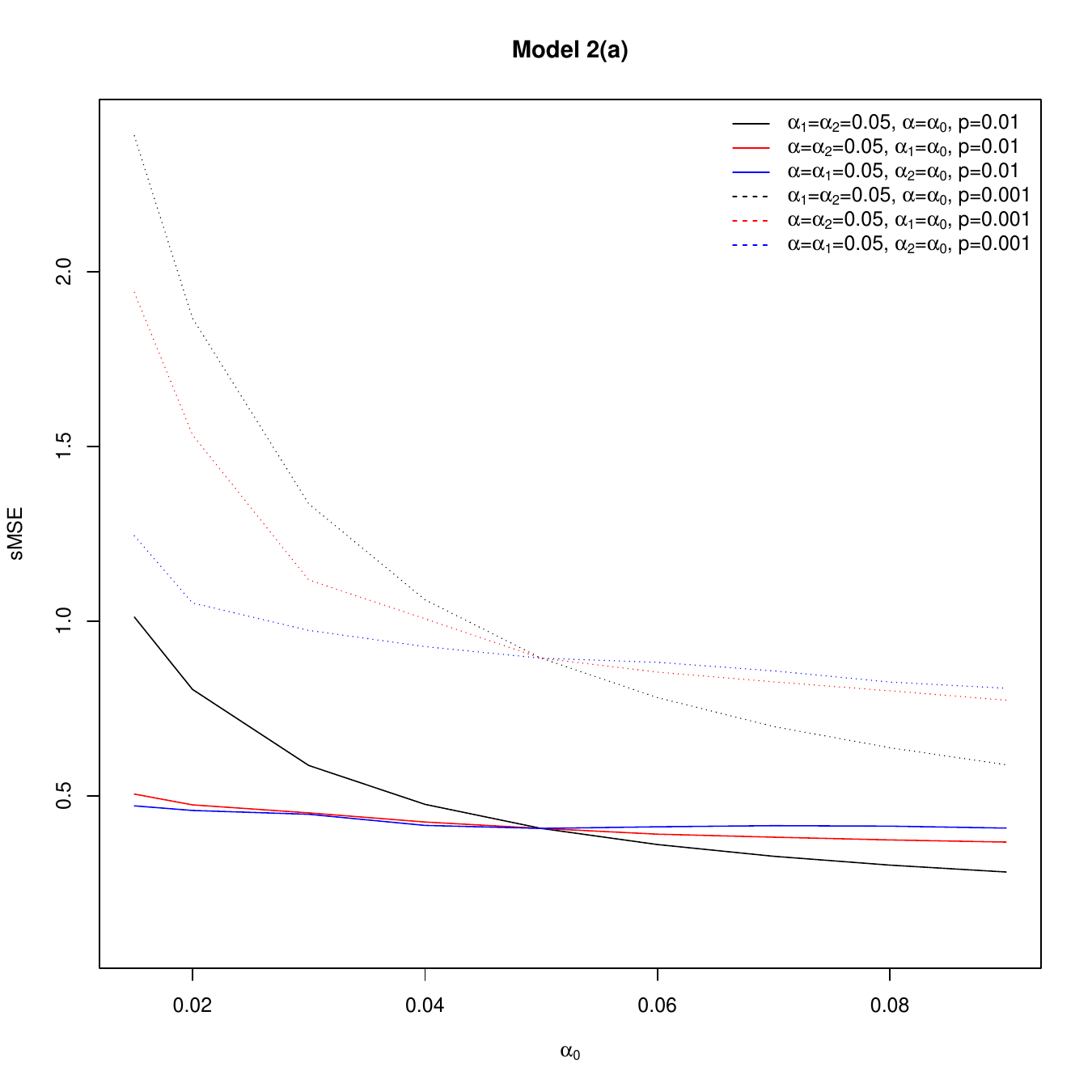}
    \end{subfigure} \caption{sMSE for different choice of intermediate levels $(\alpha, \alpha_1, \alpha_2)$.}
    \label{fg1}
\end{figure}

Let $\alpha=k/n, \alpha_1=k_1/n,\alpha_2=k_2/n.$ Figure \ref{fg1} shows the results for the five models, where the solid lines denote the results for $p=0.01$ and the dotted lines denote the results for $p=0.001$. For each curve, we fix the paramaters values of $\alpha$'s to be 0.05 and let the remaining $\alpha$ vary. Figure \ref{fg1} suggests that sMSE is rather stable for a wide range of $\alpha_1$ and $\alpha_2$. 

In order to assess the finite sample performance of our estimator, the ratio $\hat{\theta}_p/\operatorname{T G}_p(X ; Y)$ is calculated and the corresponding means and standard errors are reported in Table \ref{tb2}. Here we compare our proposed method (denoted by AIE) and the method in \cite{hou2021extreme} (denoted by HW). We set $\alpha=0.09, \alpha_1=\alpha_2=0.05.$ Table \ref{tb2} shows that the means of the ratios for AIE are closer to one than that of the ratios for HW; the standard errors of the ratios for AIE are smaller than that of the ratios for HW. Both indicate the accuracy of our proposed estimators under asymptotic independence. It is worth noting that AIE tends to underestimate by a small margin while HW tends to overstimate by a larger margin. This is due to the fact that in AIE the $\hat{\eta}$ is usually smaller than the true $\eta$. In HW, on the other hand, $\eta$ is taken as 1 by default. For both AIE and HW, the estimators at extreme level $p=0.01$ perform better than that at $p=0.001$. Model 2 exhibits the poorest performance in terms of both the means of the ratios and the standard errors, which possibly stems from the fact that Assumption \ref{asm2} is not satisfied for Model 2.
\begin{table}[] \centering
    \caption{Means of the ratios of the proposed estimators for the tail Gini functional and the true values for $p = 0.01, 0.001,$ are reported with corresponding standard deviation given in the brackets.}
    \begin{tabular}{llll}
    \hline
               &           & AIE             & HW             \\ \hline
    Model 1(a) & $p=0.01$  & 0.9263(0.3831)  & 1.3955(0.4291) \\  
               & $p=0.001$ & 0.8661(0.4416)  & 1.9696(0.6133) \\ \hline
    Model 1(b) & $p=0.01$  & 0.9028(0.3503)  & 1.3092(0.3940) \\ 
               & $p=0.001$ & 0.8583(0.4527) & 1.7911(0.6174) \\ \hline
    Model 1(c) & $p=0.01$  & 0.9137(0.5278)  & 1.4568(0.7272) \\ 
               & $p=0.001$ & 0.7995(0.5506)  & 2.0634(1.0907) \\ \hline
    Model 1(d) & $p=0.01$  & 0.9528(0.4914)  & 1.6627(0.7273) \\ 
               & $p=0.001$ & 0.9641(0.6230)  & 2.9681(1.3097) \\ \hline
    Model 2 & $p=0.01$  & 0.9541(3.7531)  & 1.2149(4.4512) \\ 
               & $p=0.001$ & 0.8536(0.9595)  & 1.3889(1.5865) \\ \hline
    \end{tabular}

        \label{tb2}
    \end{table}

\begin{figure}
        \centering
        \begin{subfigure}[b]{0.18\textwidth}
            \centering
            \includegraphics[width=\linewidth]{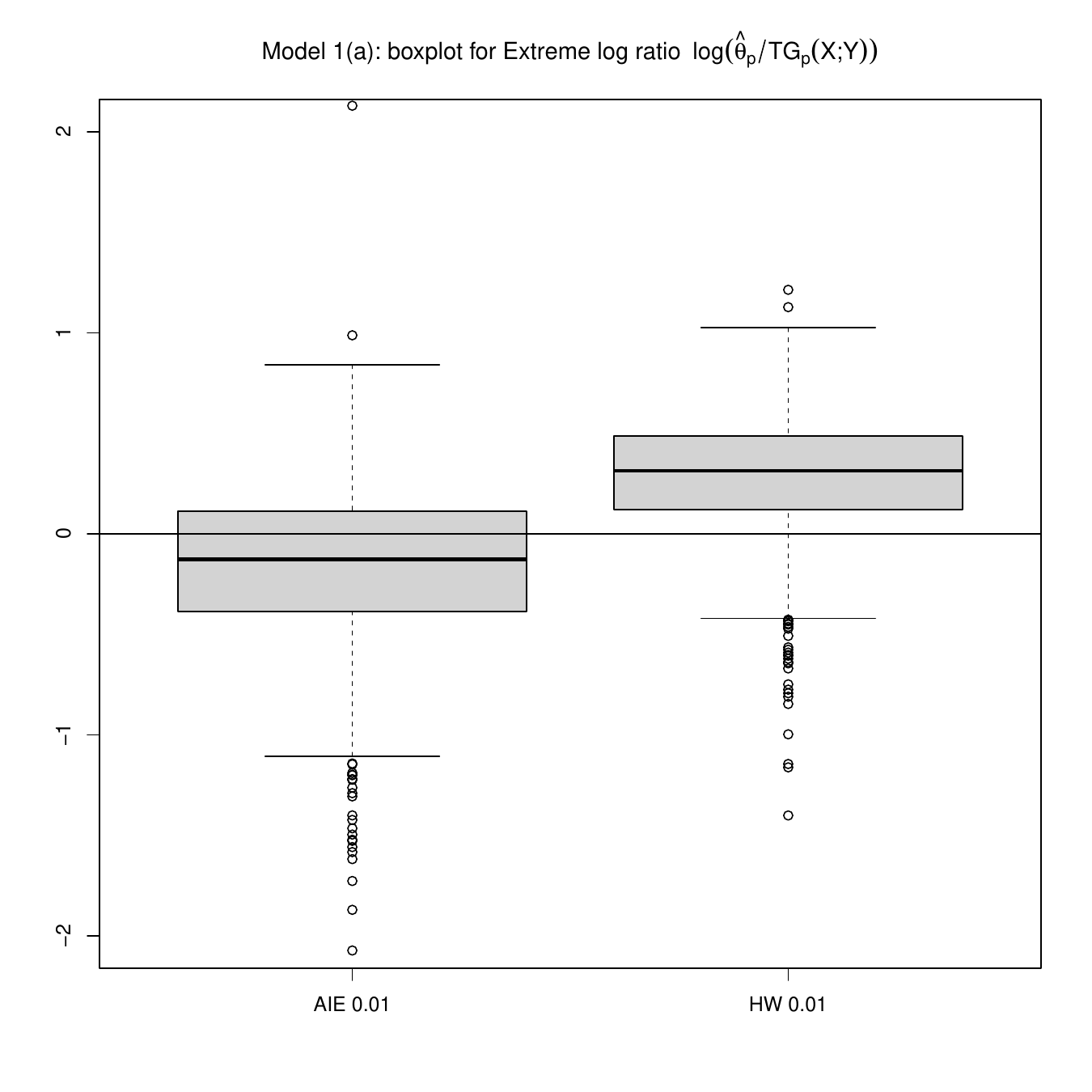}
        \end{subfigure}
        \hfill
        \begin{subfigure}[b]{0.18\textwidth}
            \centering
            \includegraphics[width=\linewidth]{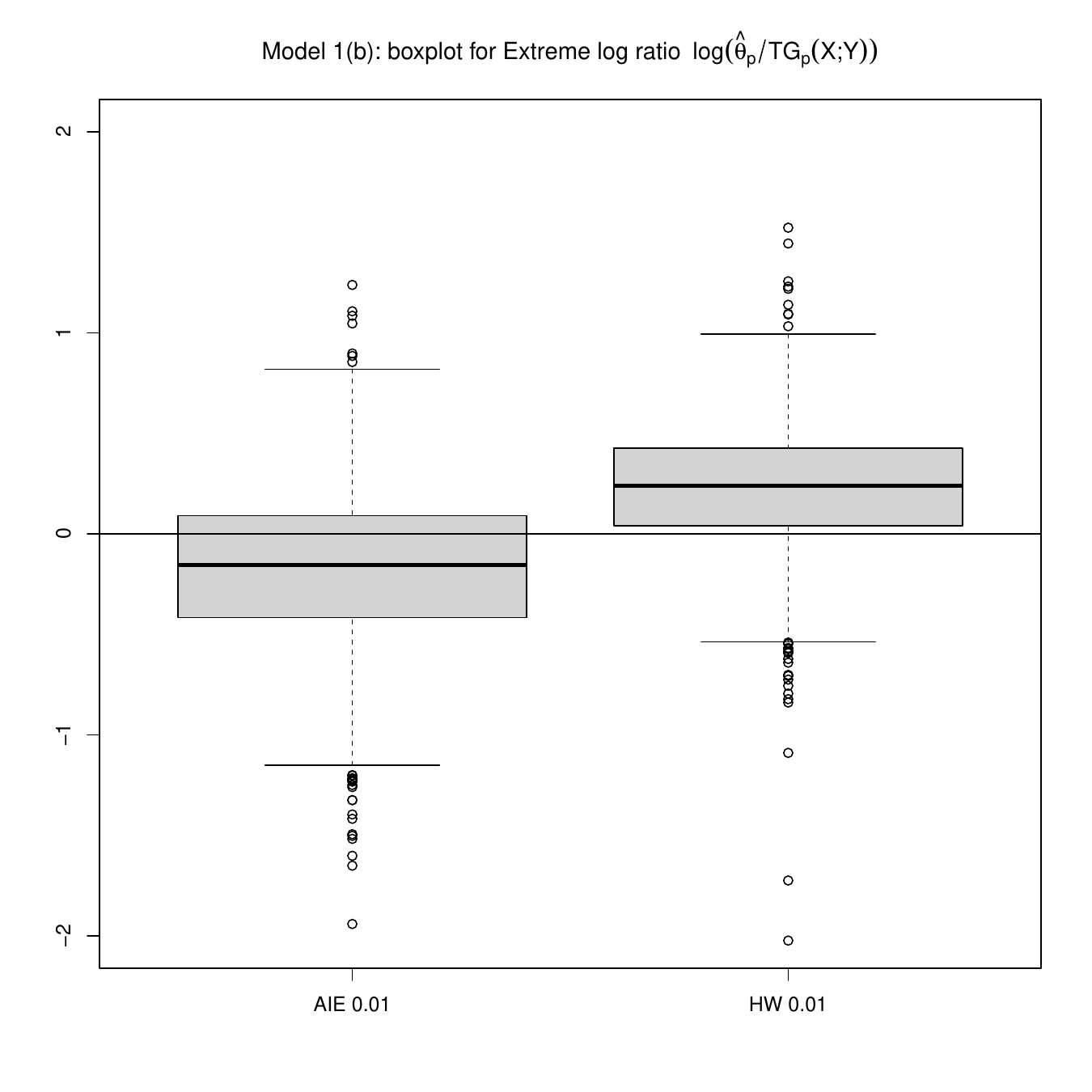}
        \end{subfigure}
        \hfill
         \begin{subfigure}[b]{0.18\textwidth}
            \centering
             \includegraphics[width=\linewidth]{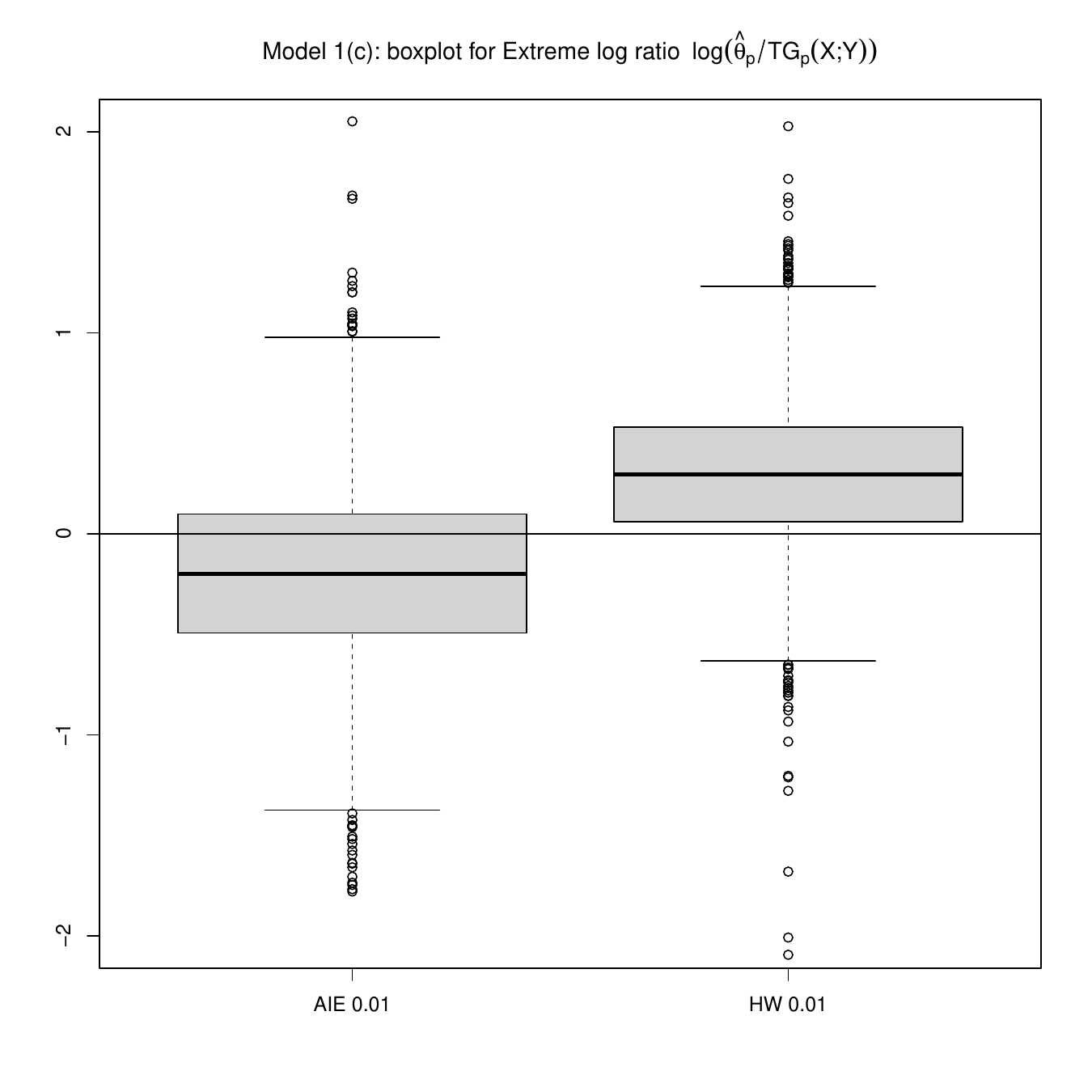}
         \end{subfigure}\hfill
        \begin{subfigure}[b]{0.18\textwidth}
            \centering
            \includegraphics[width=\linewidth]{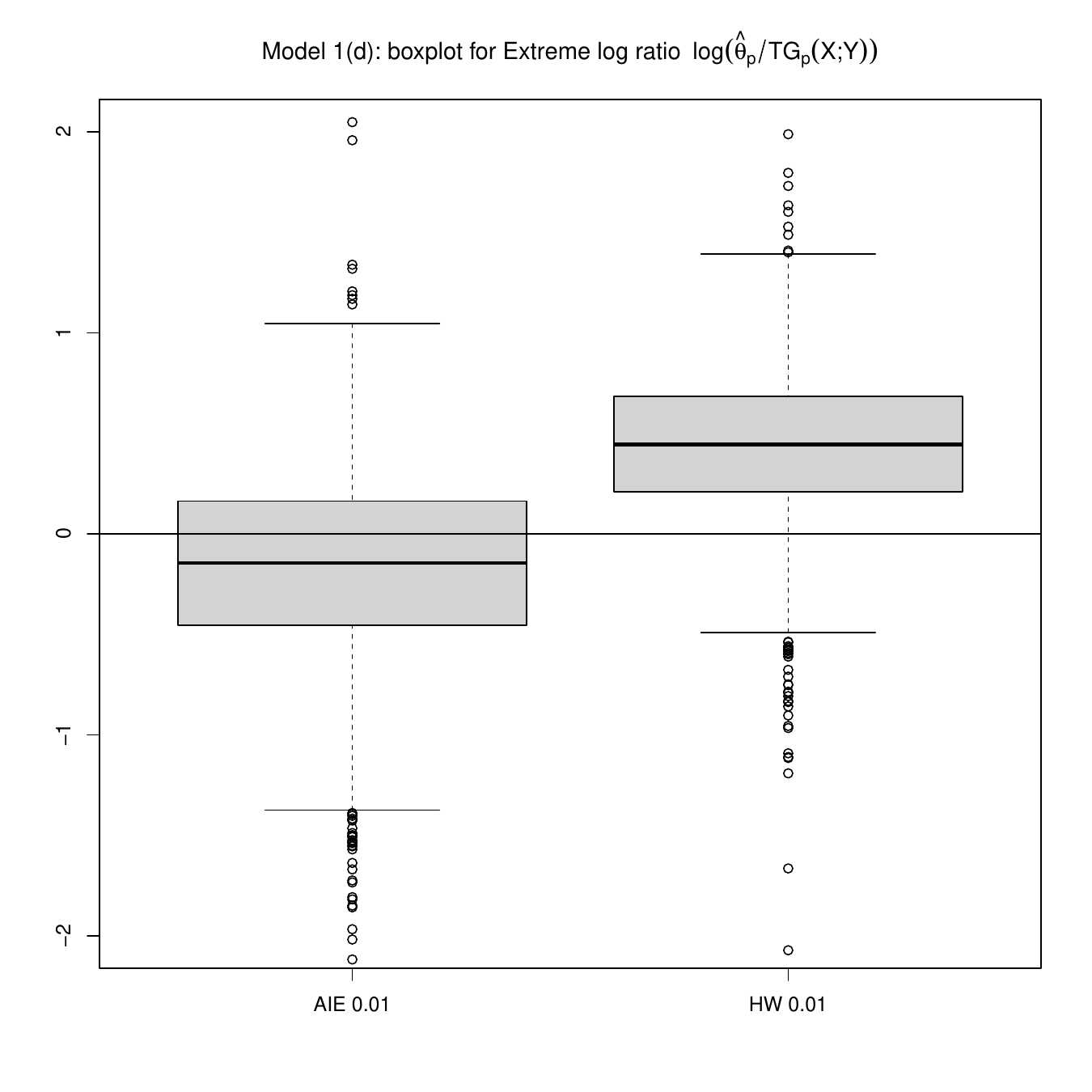}
        \end{subfigure}
        \hfill
        \begin{subfigure}[b]{0.18\textwidth}
            \centering
            \includegraphics[width=\linewidth]{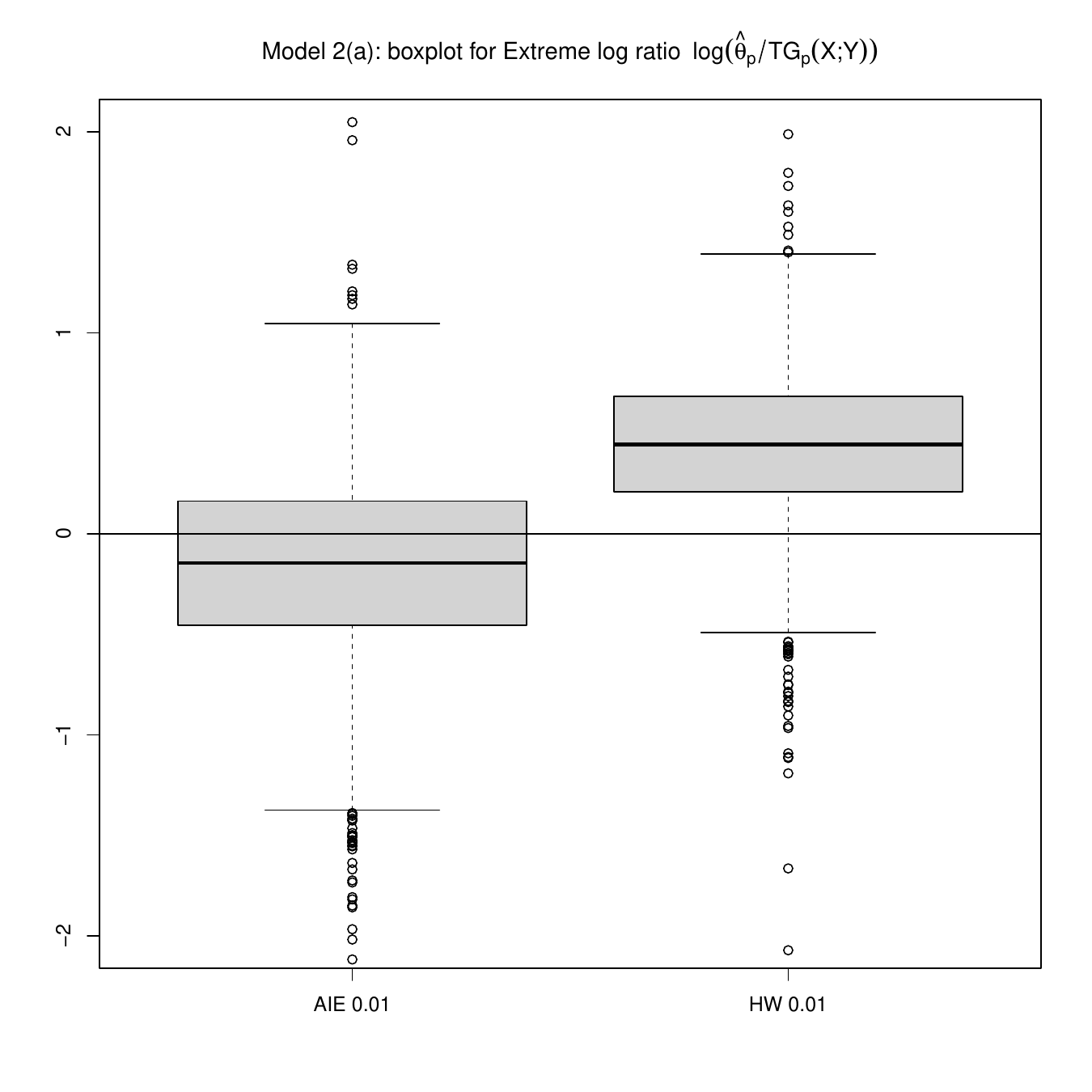}
        \end{subfigure} 
        \hfill
        \begin{subfigure}[b]{0.18\textwidth}
            \centering
            \includegraphics[width=\linewidth]{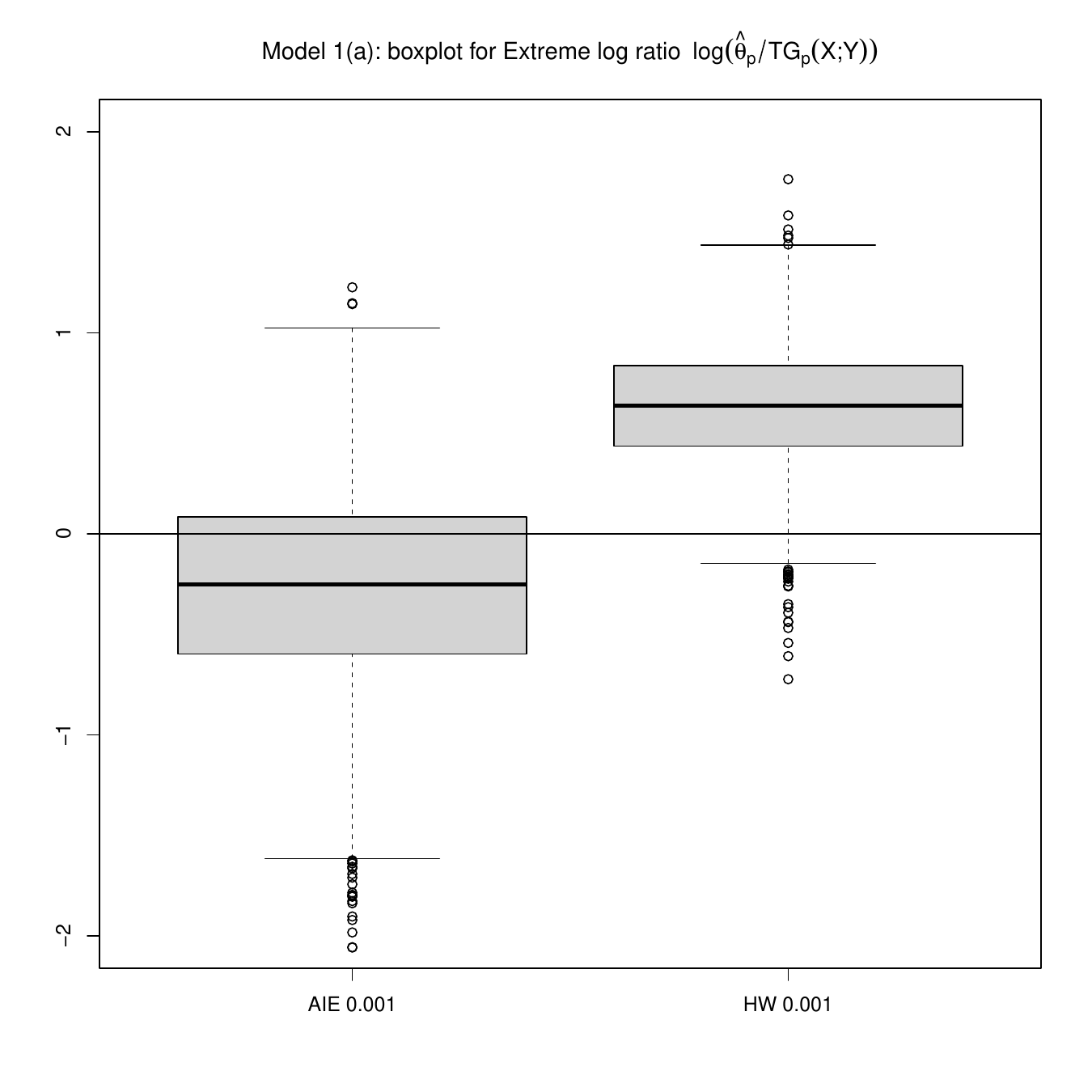}
        \end{subfigure}
        \hfill
        \begin{subfigure}[b]{0.18\textwidth}
            \centering
            \includegraphics[width=\linewidth]{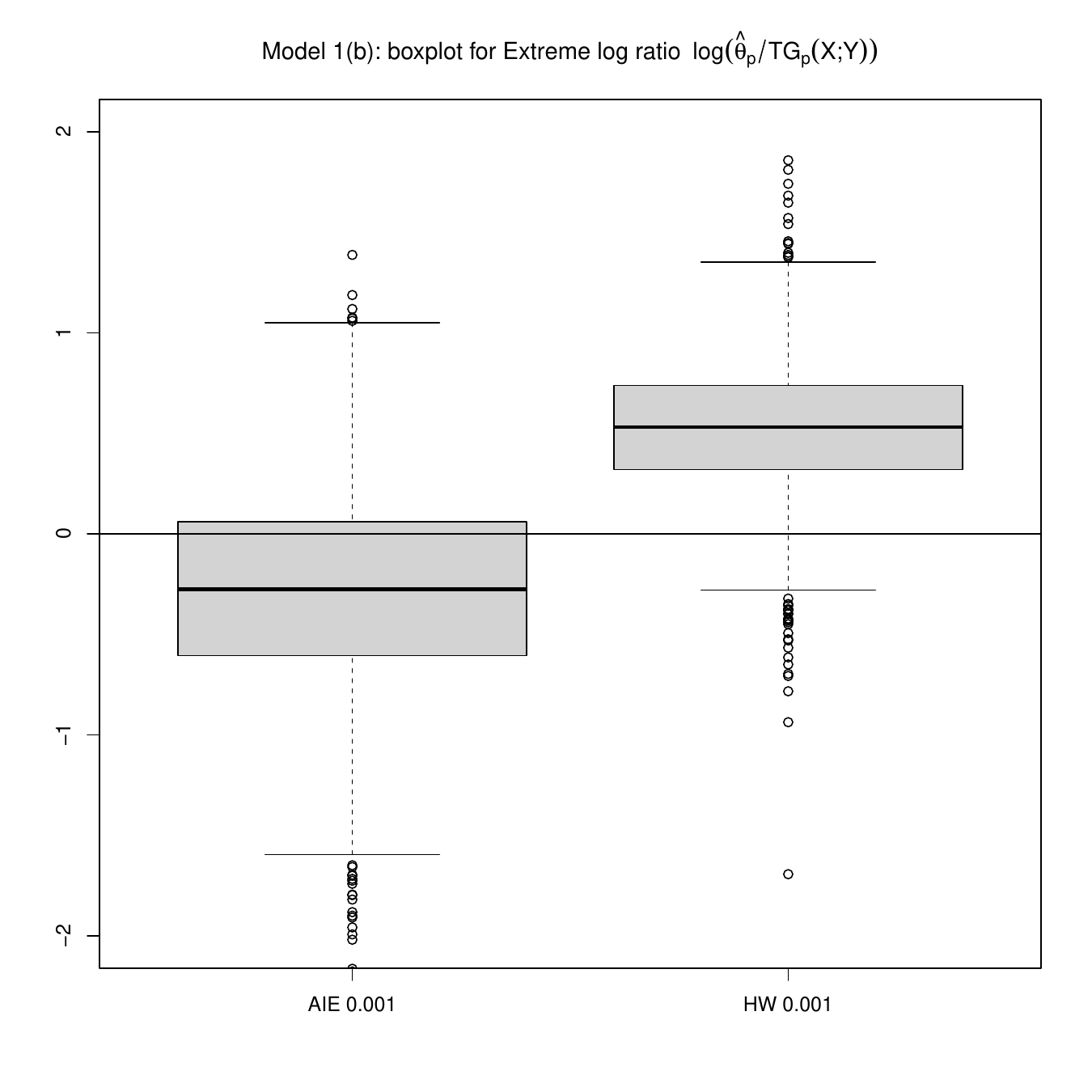}
        \end{subfigure}\hfill
        \begin{subfigure}[b]{0.18\textwidth}
            \centering
            \includegraphics[width=\linewidth]{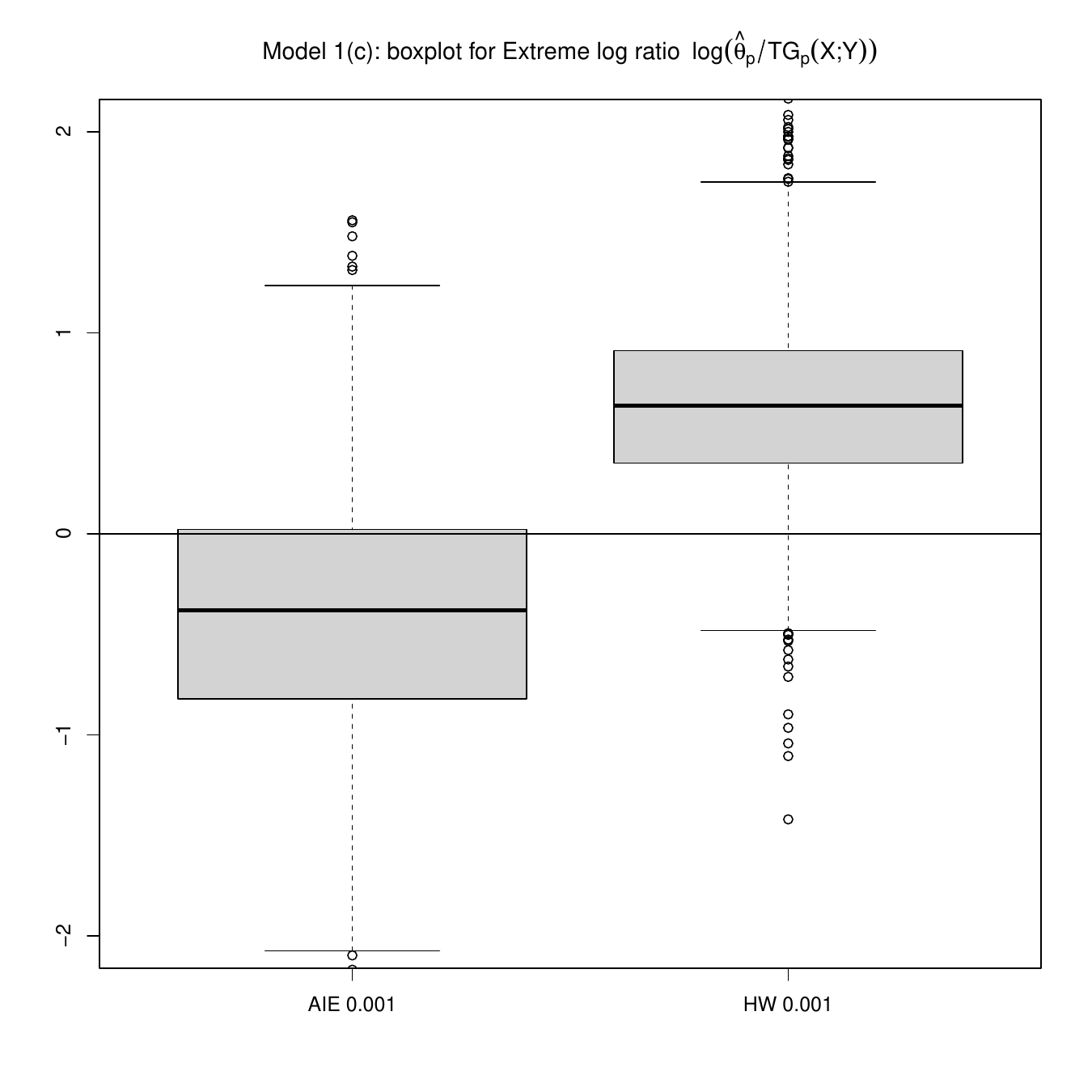}
        \end{subfigure}
        \hfill
        \begin{subfigure}[b]{0.18\textwidth}
            \centering
            \includegraphics[width=\linewidth]{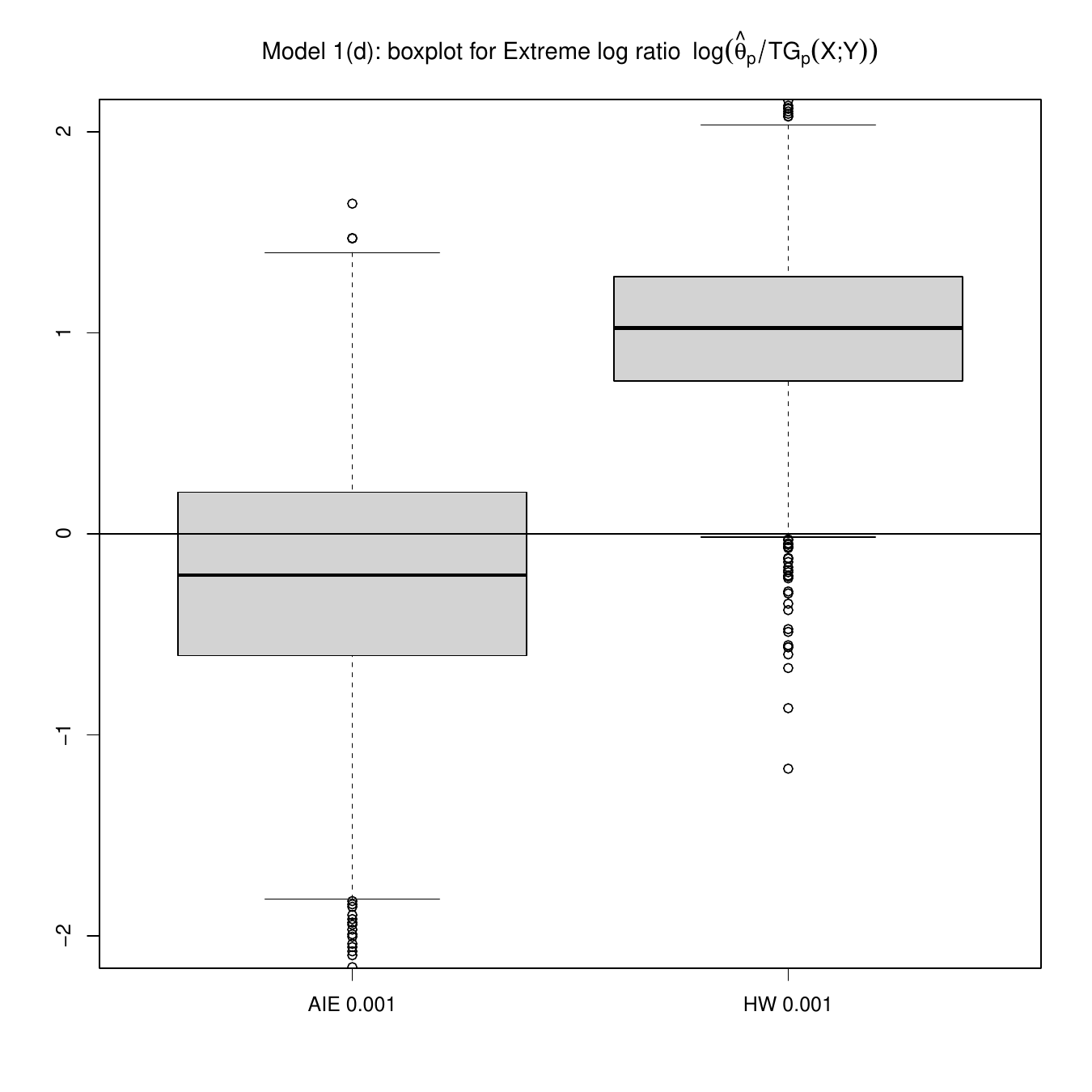}
        \end{subfigure}\hfill
        \begin{subfigure}[b]{0.18\textwidth}
            \centering
            \includegraphics[width=\linewidth]{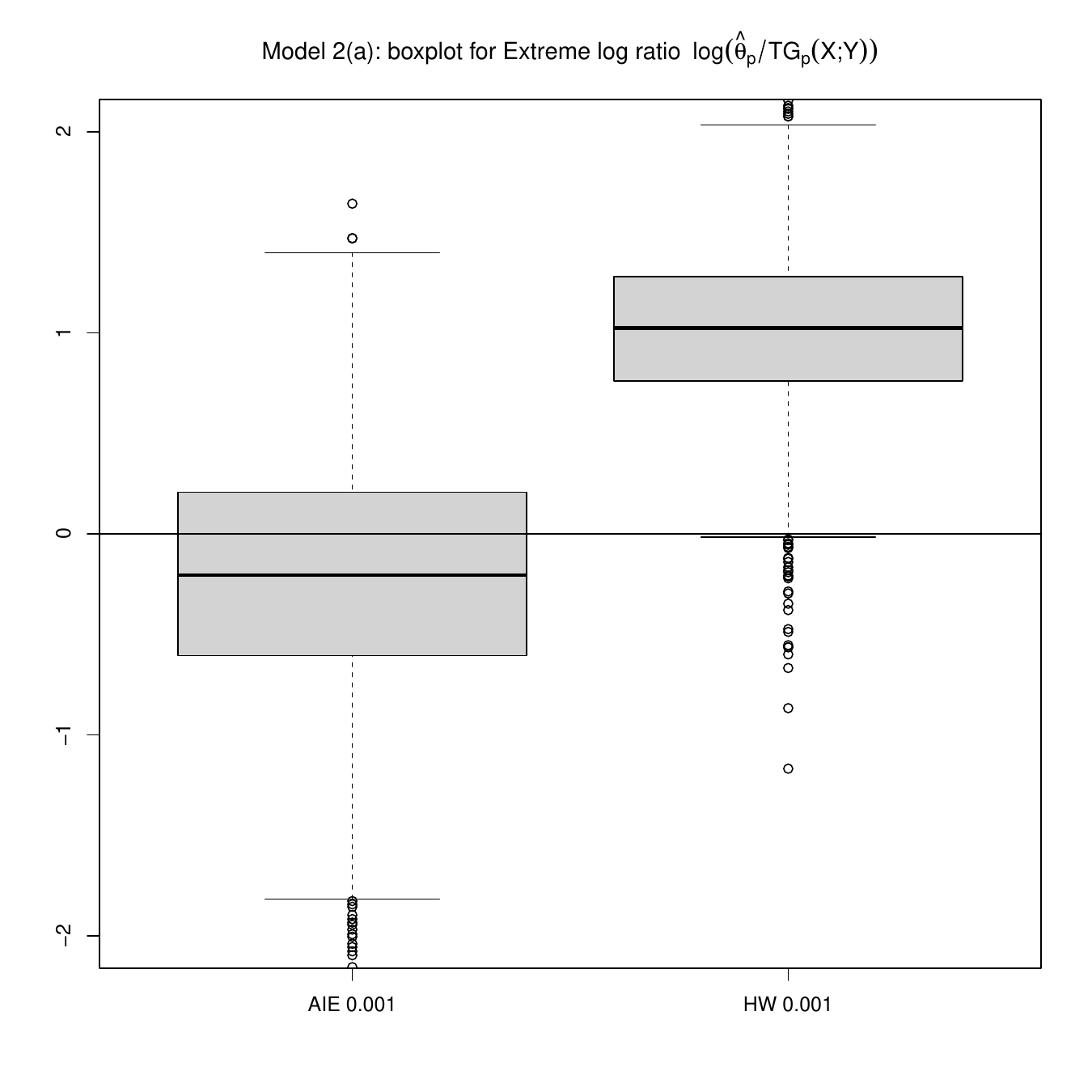}
        \end{subfigure}
           \caption{Boxplots of log ratios with $p = 0.01$ (upper) and $p = 0.001$ (bottom).}
           \label{fg2}
    \end{figure}

    In addition, we present the boxplots of $\log \left(\hat{\theta}_{k/n}/\mathrm{TG}_{k / n}(X ; Y)\right)$. The boxplots depicted in Figure \ref{fg2} illustrate a symmetrical distribution of the estimated values obtained through our proposed estimator, predominantly centered around one. In contrast, the estimator in \cite{hou2021extreme} performs poorly because it fails to take $\eta$ into consideration and overestimats the tail Gini functional.

\begin{figure}
    \centering
    \begin{subfigure}[b]{0.18\textwidth}\centering\includegraphics[width=\linewidth]{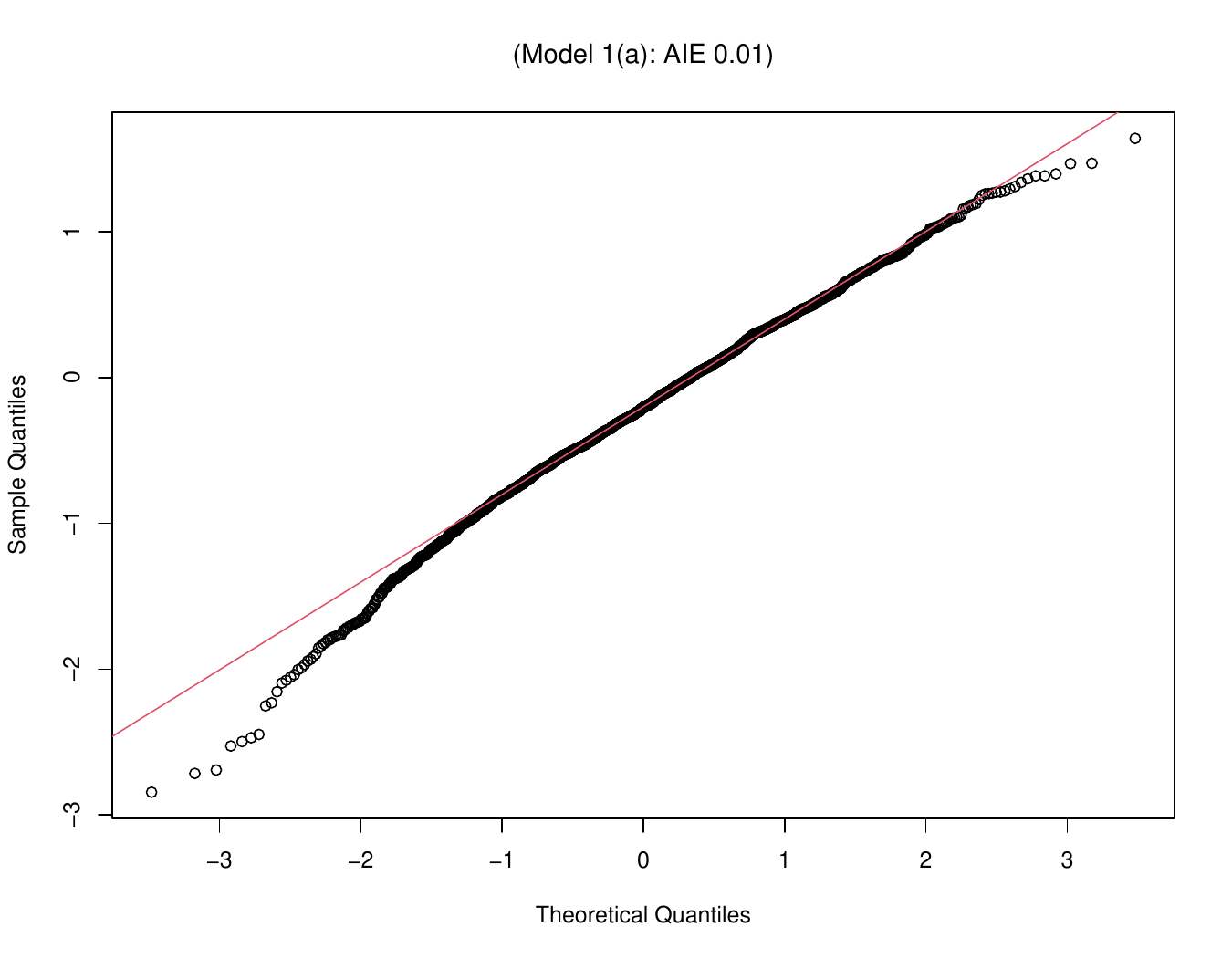}
   \end{subfigure}
   \hfill
   \begin{subfigure}[b]{0.18\textwidth}
       \centering
       \includegraphics[width=\linewidth]{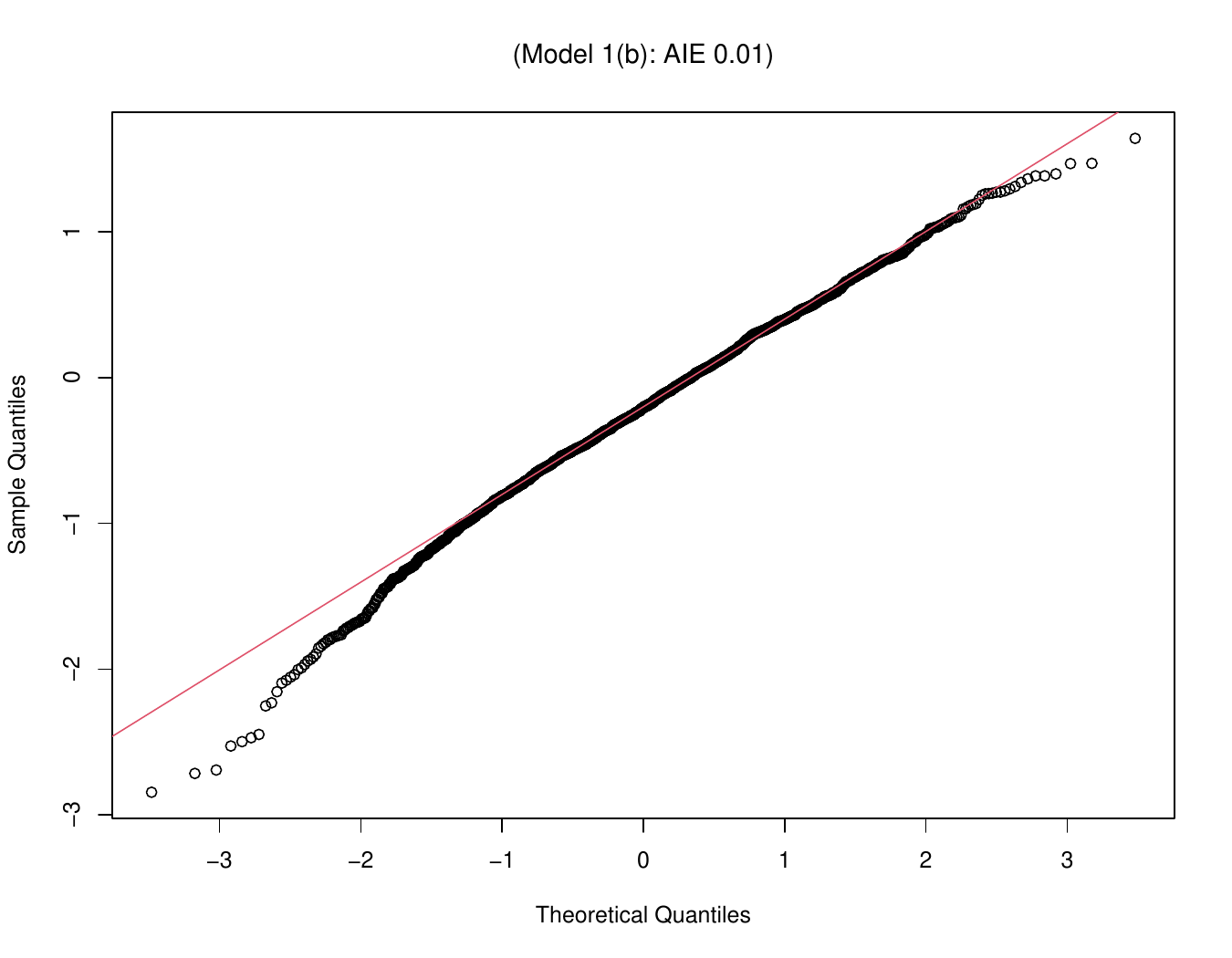}
   \end{subfigure}
   \hfill
   \begin{subfigure}[b]{0.18\textwidth}
       \centering
       \includegraphics[width=\linewidth]{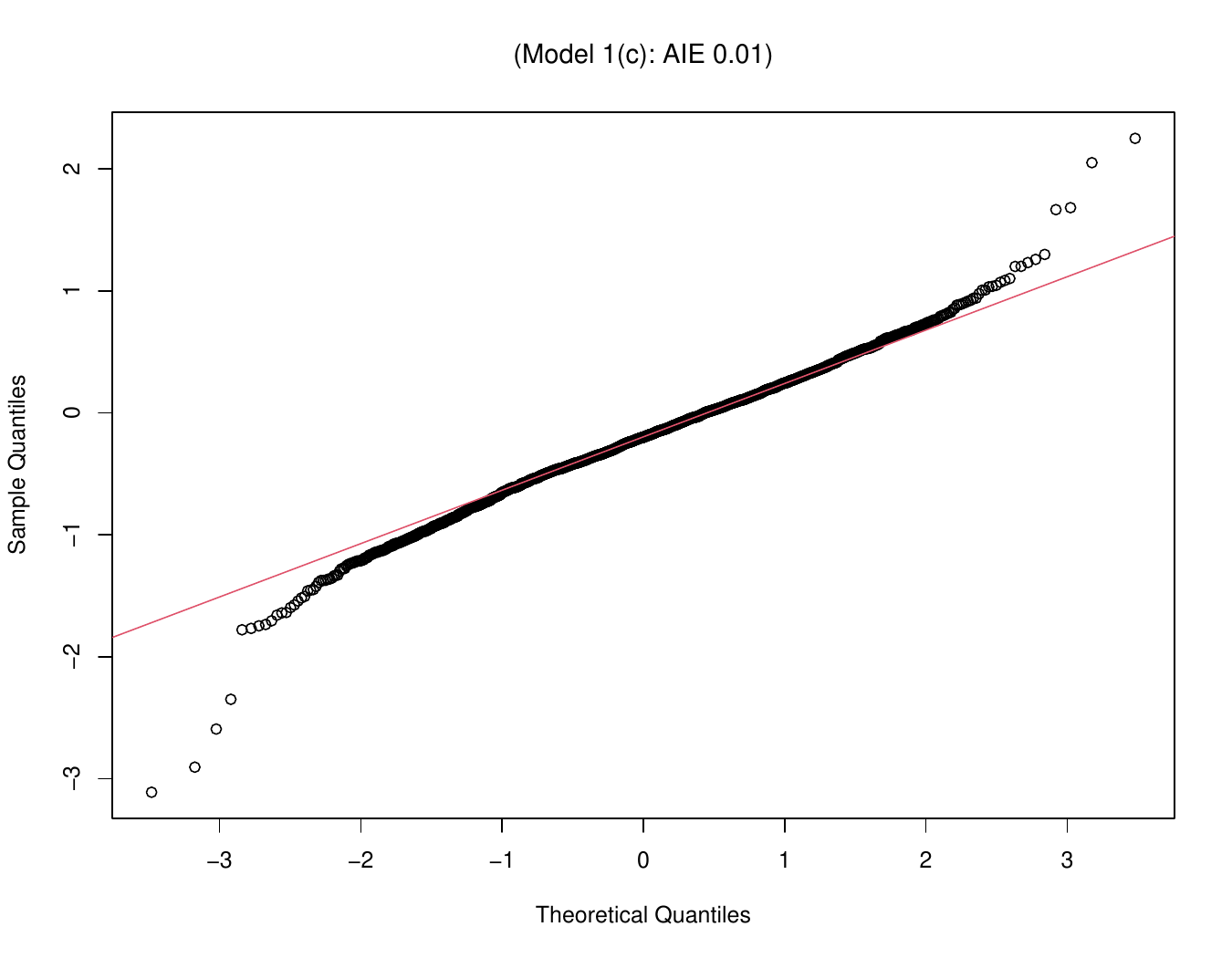}
   \end{subfigure}\hfill
   \begin{subfigure}[b]{0.18\textwidth}
       \centering
       \includegraphics[width=\linewidth]{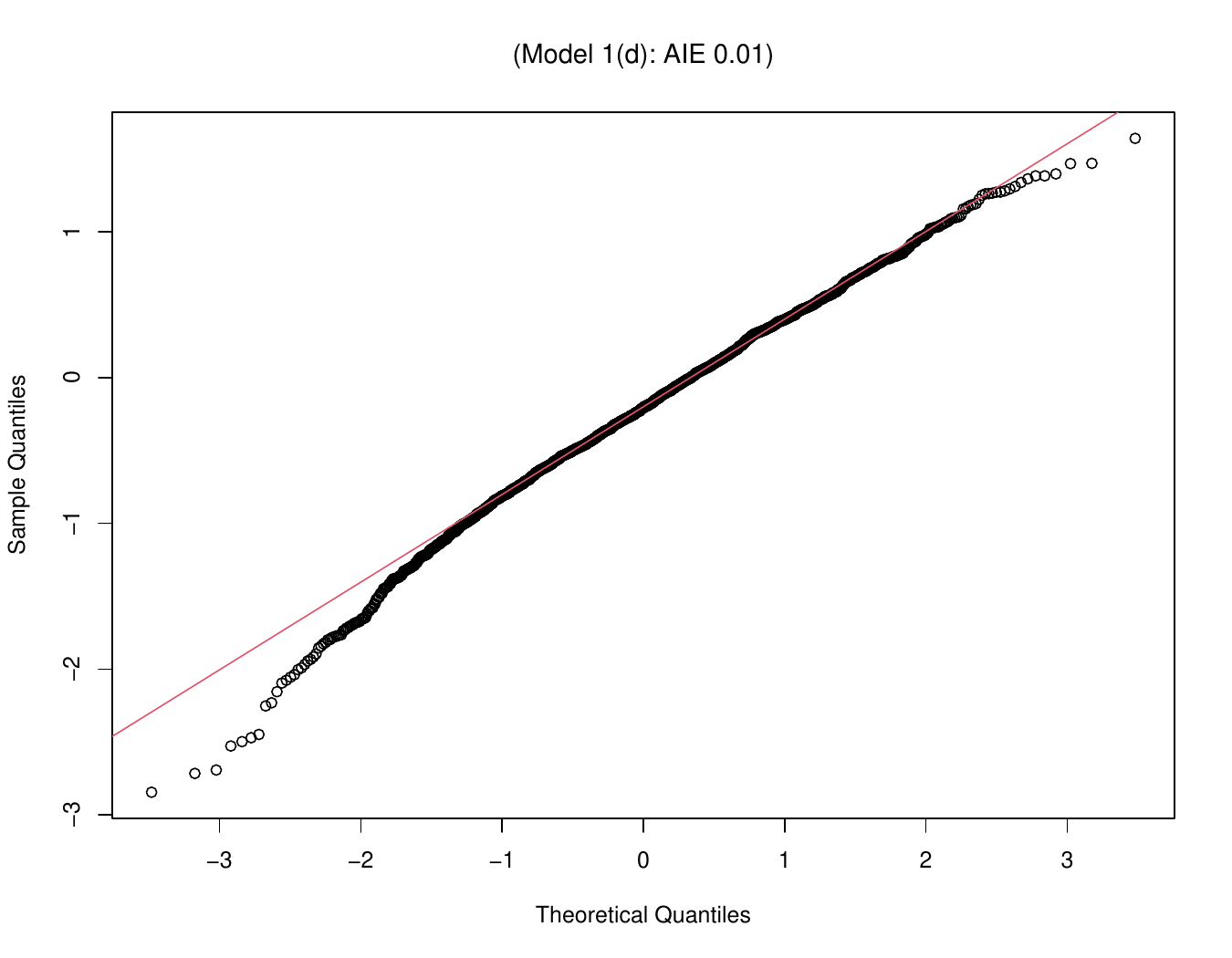}
   \end{subfigure}
   \hfill
   \begin{subfigure}[b]{0.18\textwidth}
       \centering
       \includegraphics[width=\linewidth]{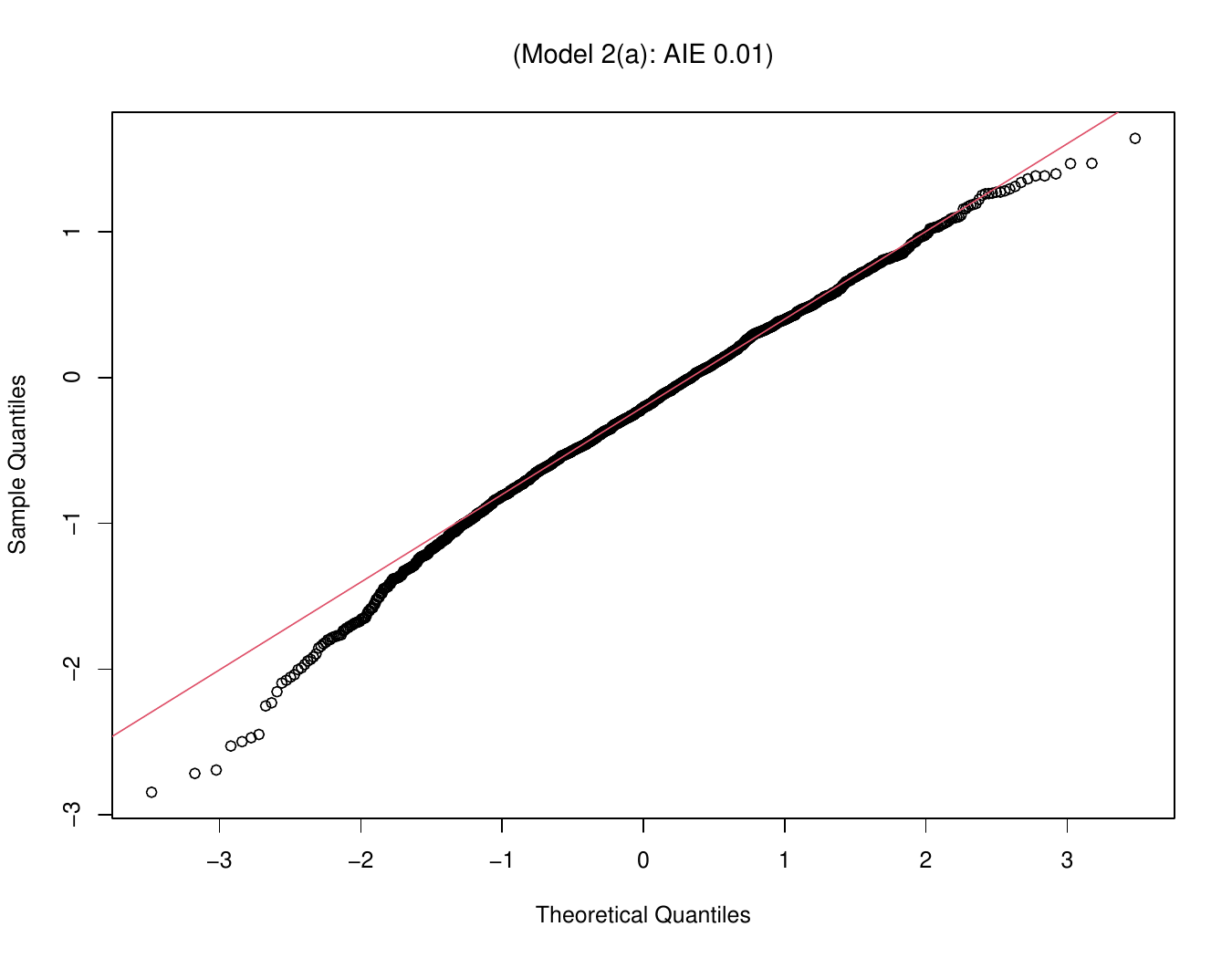}\end{subfigure}\hfill
   \begin{subfigure}[b]{0.18\textwidth}
       \centering
       \includegraphics[width=\linewidth]{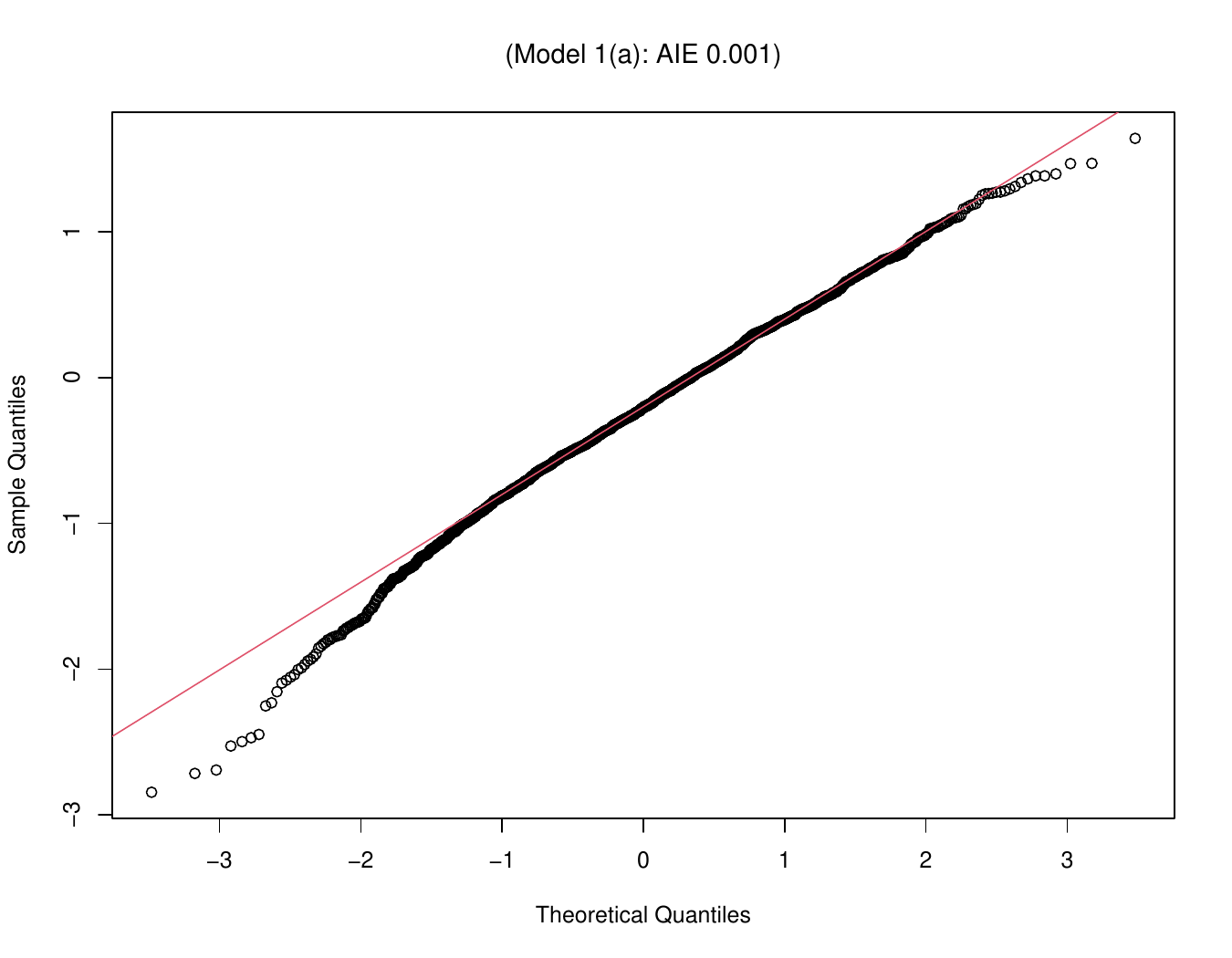}
   \end{subfigure}
   \hfill
   \begin{subfigure}[b]{0.18\textwidth}
       \centering
       \includegraphics[width=\linewidth]{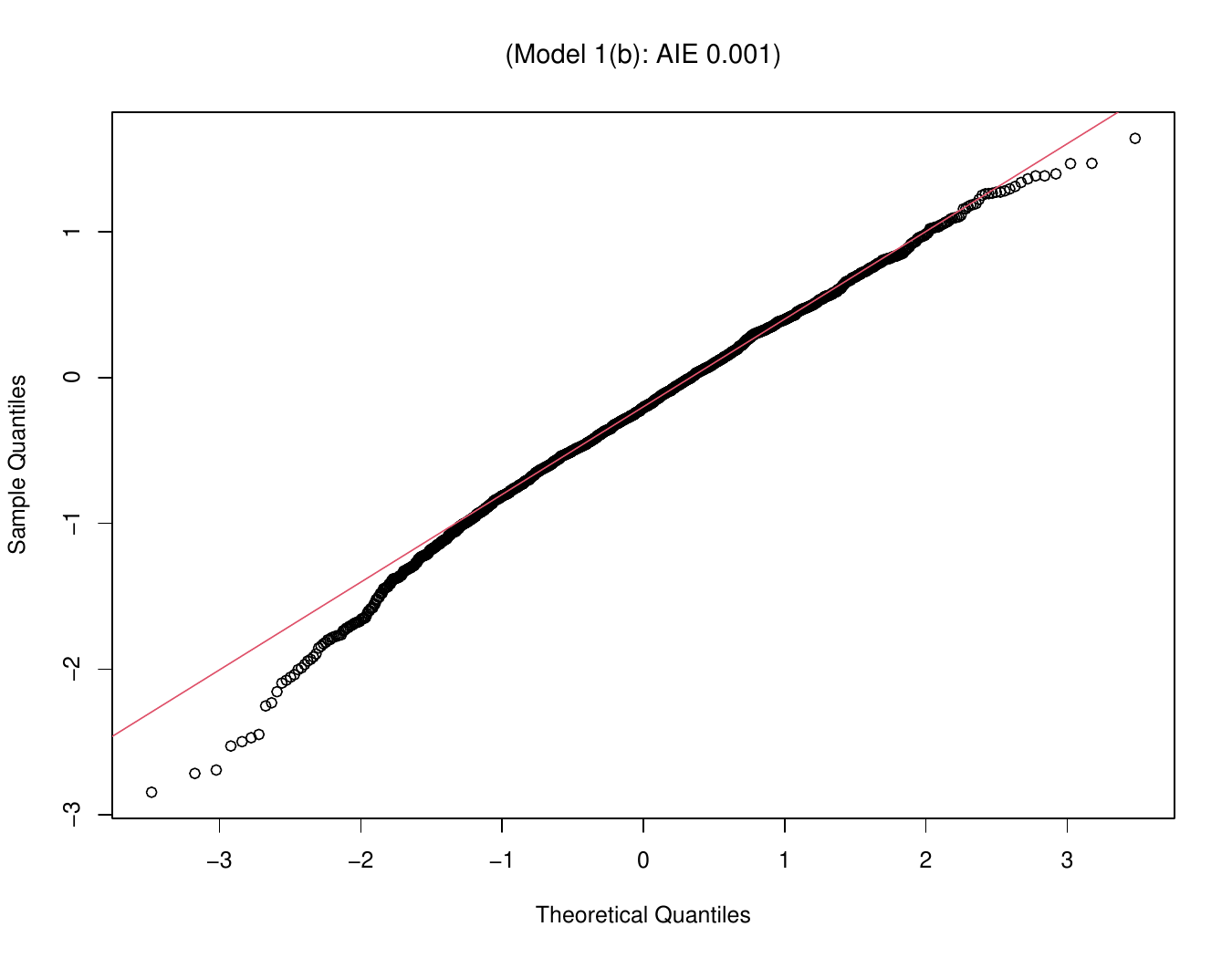}
   \end{subfigure}\hfill
   \begin{subfigure}[b]{0.18\textwidth}
       \centering
       \includegraphics[width=\linewidth]{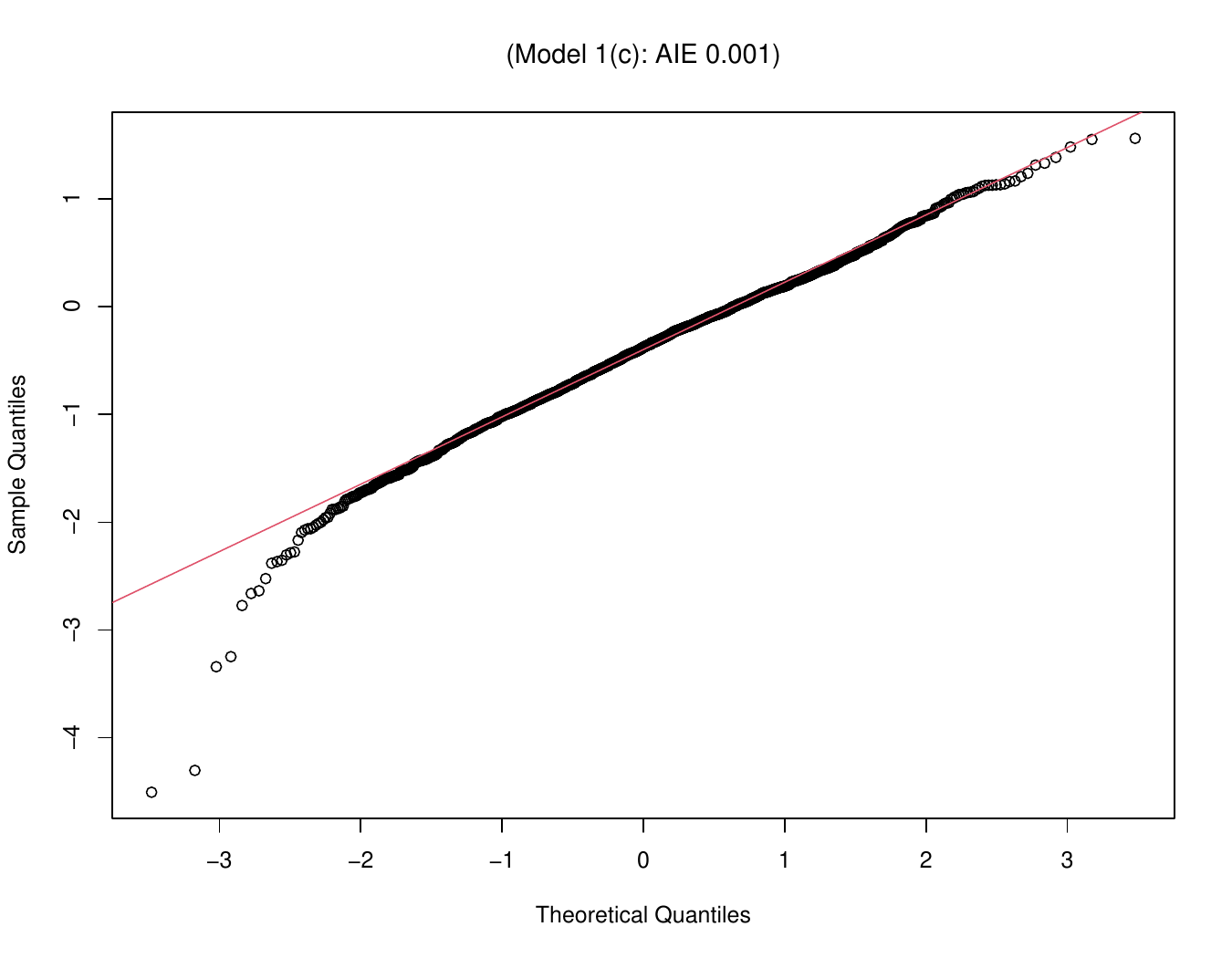}
   \end{subfigure}
   \hfill
   \begin{subfigure}[b]{0.18\textwidth}
       \centering
       \includegraphics[width=\linewidth]{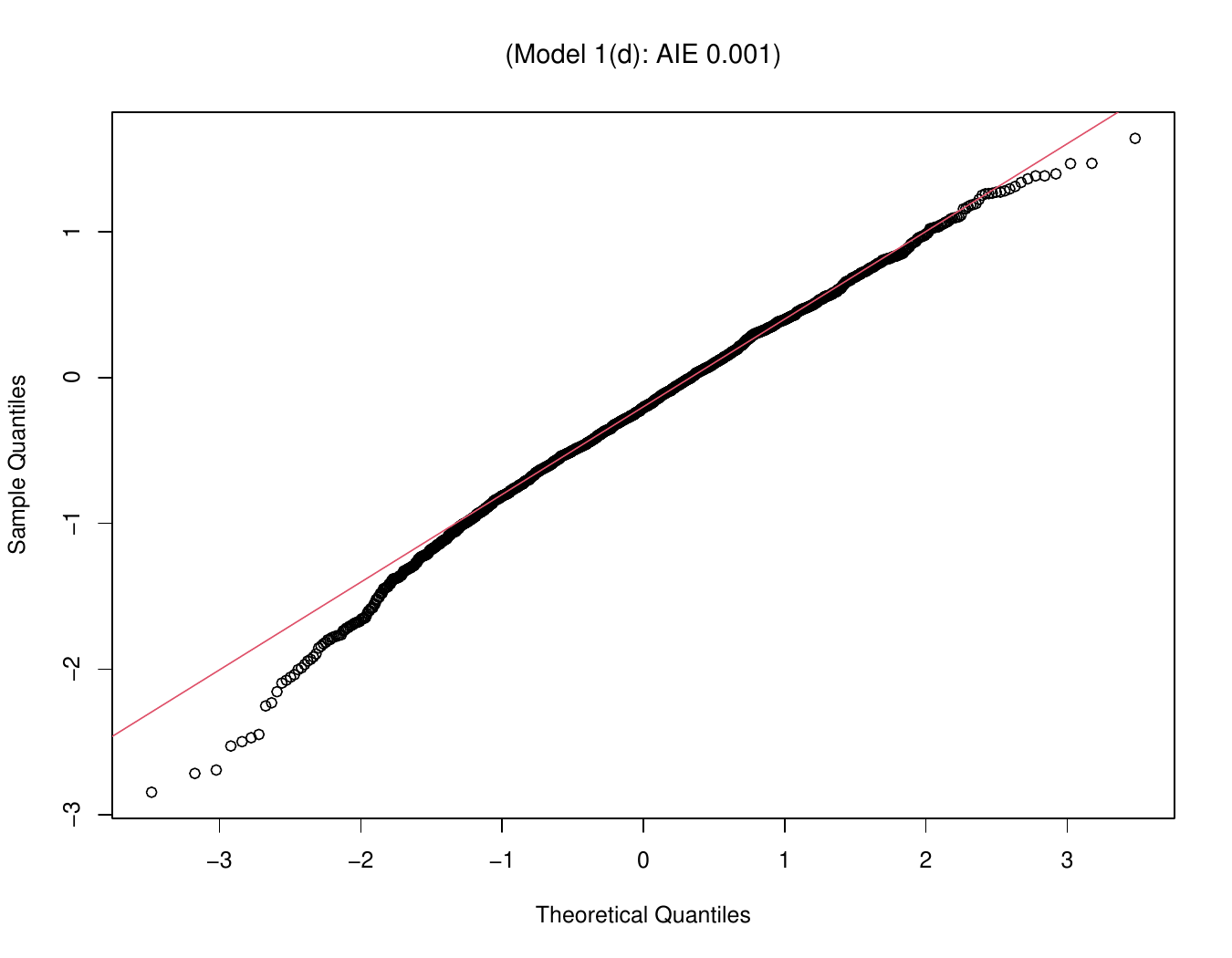}
   \end{subfigure}\hfill
   \begin{subfigure}[b]{0.18\textwidth}
       \centering
       \includegraphics[width=\linewidth]{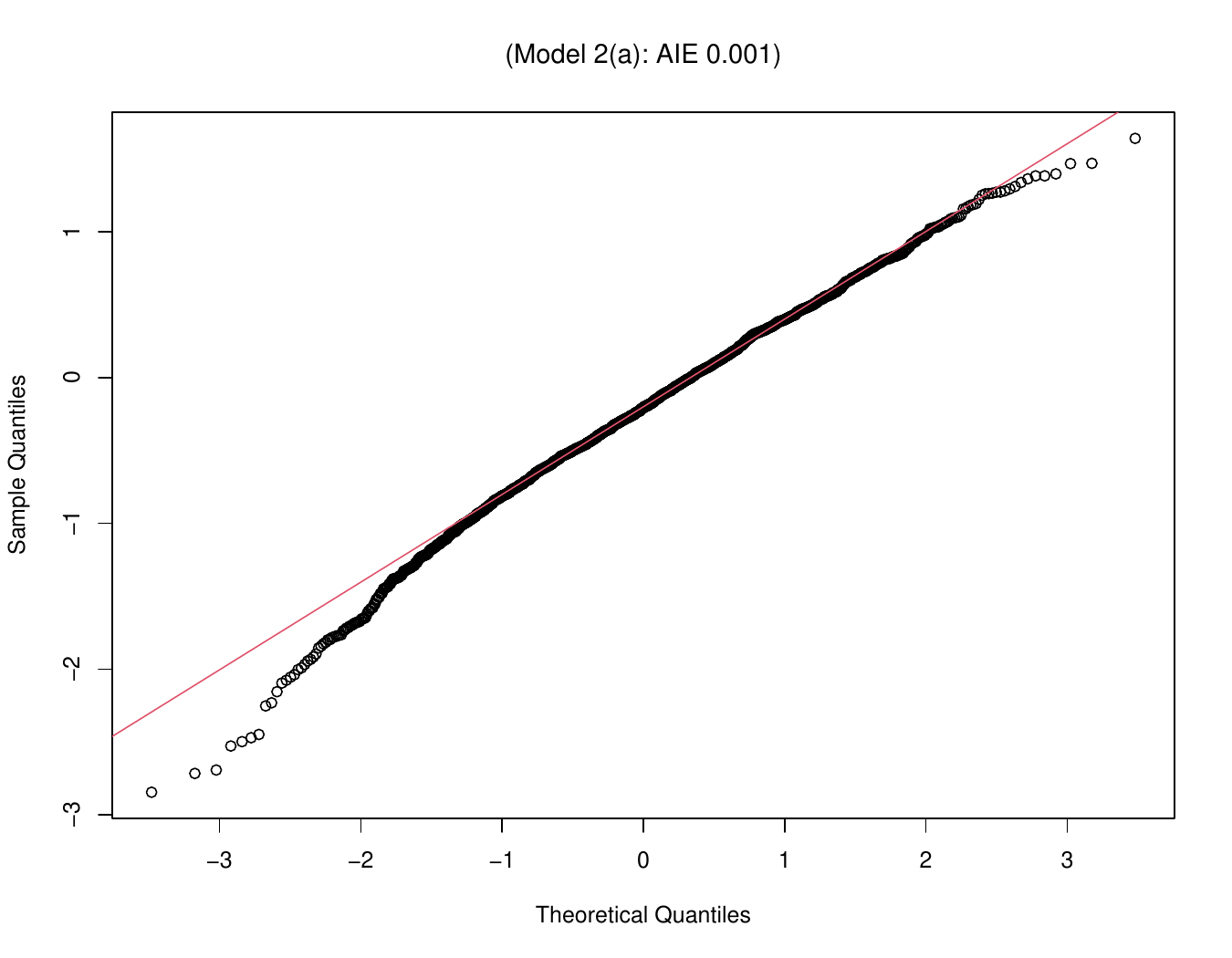}
   \end{subfigure}\hfill
    \caption{QQ plots of log ratios for AIE with $p = 0.01$ (upper) and $p = 0.001$ (bottom).}
    \label{fg3}
\end{figure}

   To show the asymptotic property of our proposed estimators, we also compare the sample quantiles of log-ratios at all levels with the normal quantiles by using QQ plots. Figure \ref{fg3} shows that most of the scatters line up on the red straight line, which indicates no big difference from a normal distribution.

\section{Application}\label{sec4}

In this section, we employ our estimator for the tail Gini functional on a dataset encompassing the daily stock price of the Hang Seng Index (HSI) and 17 companies from December 20, 1992 to December 30, 2022. Our aim is to examine the impact on individual stocks during the occurrence of an exceptionally high-risk event associated with the systemic variable, which is HSI in this application. 

\begin{figure}
    \centering
    \includegraphics[width=1\textwidth]{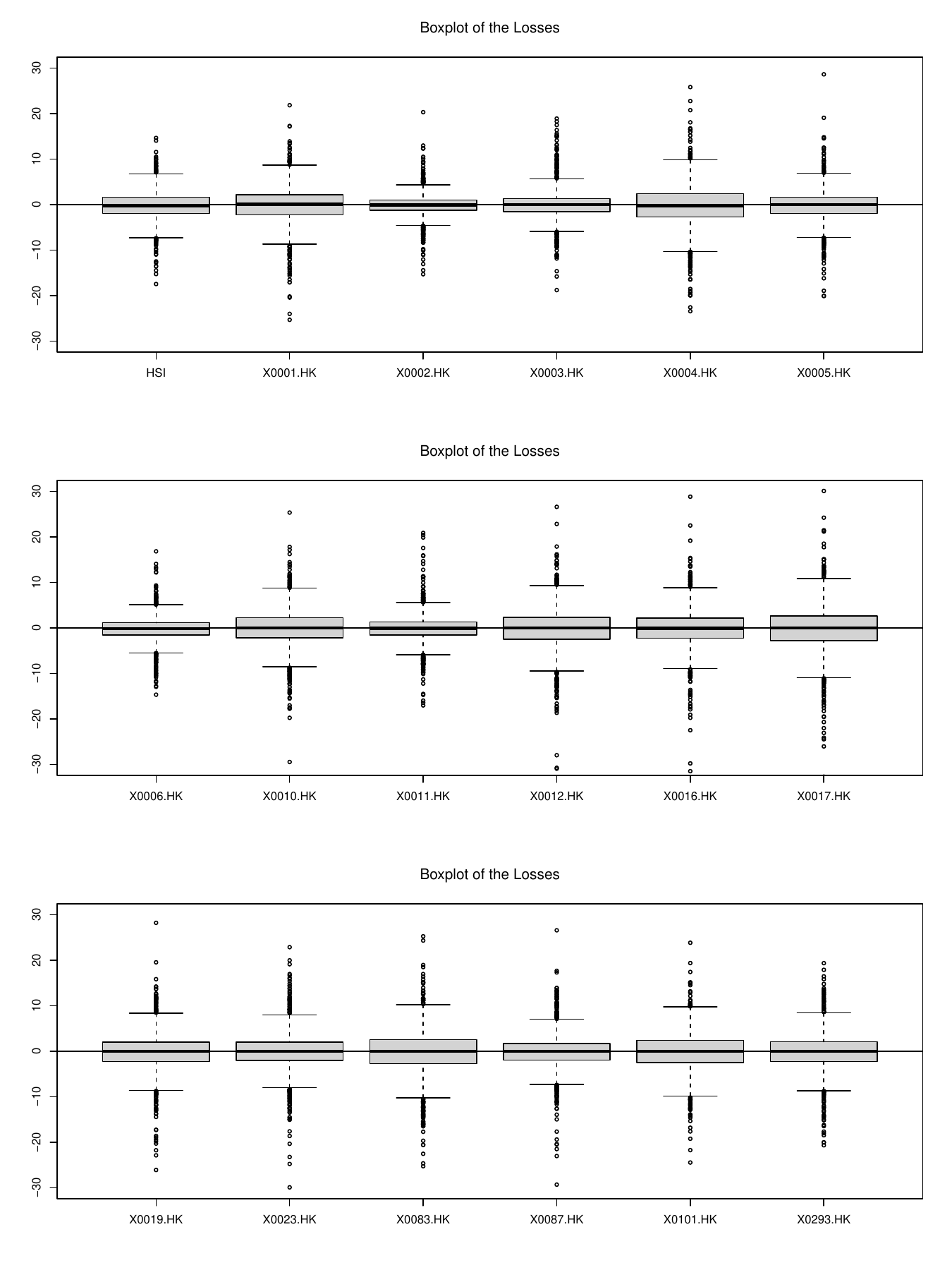}\hfill
 \caption{Boxplots of the losses.}
 \label{fg4}
\end{figure}

\begin{table}[]
    \caption{Summary statistics, TQCC, p-values and estimated values.}
    \centering
    \scalebox{0.9}{

        \begin{tabular}{llllllllll}
        \hline
        Ticker   & Firms           & Mean (\%) & SD (\%) & TQCC   & $p$-value & $\hat{\gamma}_1$ & $\hat{\eta}$ & $\hat{\theta}_{0.01}$ & $\hat{\theta}_{0.001}$ \\ \hline
        HSI      & Heng Seng Index & -0.1647   & 3.1101  &     --   &    --     &         --         &      --        &         --              &       --       \\
        X0001.HK & CKH Holdings    & -0.1482   & 4.2117  & 0.0001 & 0.9997  & 0.3744           & 0.8597       & 3.4121                & 5.5498                 \\
        X0002.HK & CLP Holdings    & -0.1278   & 2.5857  & 0.0000 & 1.0000  & 0.4099           & 0.8214       & 1.2826                & 1.9974                 \\
        X0003.HK & HK \& China GAS & -0.0103   & 3.2390  & 0.0000 & 1.0000  & 0.5285           & 0.7501       & 1.9425                & 3.0459                 \\
        X0004.HK & Wharf Holdings  & -0.2317   & 4.9306  & 0.0000 & 1.0000  & 0.3904           & 0.8649       & 3.7878                & 6.4946                 \\
        X0005.HK & HSBC Holdings   & -0.4176   & 9.6040  & 0.1043 & 0.0000  & --          & --  &  --    & -- \\
        X0006.HK & Power Assets    & -0.1408   & 2.6953  & 0.0000 & 1.0000  & 0.4203           & 0.7591       & 0.9901                & 1.2547                 \\
        X0010.HK & Hang Lung Group & -0.0650   & 4.2654  & 0.0003 & 0.9985  & 0.3483           & 0.8590       & 2.8533                & 4.3601                 \\
        X0011.HK & Hang Seng Bank  & -0.0844   & 3.2972  & 0.0207 & 0.1063  & 0.4536           & 0.8824       & 3.4732                & 7.2628                 \\
        X0012.HK & Henderson Land  & -0.1587   & 4.6614  & 0.0014 & 0.9630  & 0.4031           & 0.8767       & 3.2079                & 5.7107                 \\
        X0016.HK & SHK PPT         & -0.1871   & 4.4221  & 0.0003 & 0.9970  & 0.3993           & 0.8210       & 2.8891                & 4.3858                 \\
        X0017.HK & New World Dev   & -0.3594   & 9.0423  & 0.3364 & 0.0000  & --           & --     &     --    &      --   \\
        X0019.HK & Swire Pacific A & -0.1326   & 4.3048  & 0.0001 & 0.9997  & 0.4527           & 0.8805       & 4.7142                & 9.7794                 \\
        X0023.HK & Bank of E Asia  & -0.0510   & 4.2934  & 0.0003 & 0.9985  & 0.4229           & 0.7898       & 2.8040                & 4.0235                 \\
        X0083.HK & Sino Land       & -0.4961   & 11.2818 & 0.0699 & 0.0000  & --         & --     &   --   &   --    \\
        X0087.HK & Swire Pacific B & -0.1126   & 3.8678  & 0.0001 & 0.9999  & 0.4538           & 0.8948       & 4.6822                & 10.1563                \\
        X0101.HK & Hang Lung PPT   & -0.0933   & 4.3302  & 0.0016 & 0.9689  & 0.3646           & 0.8846       & 3.2072                & 5.4987                 \\
        X0293.HK & Cathay Pacific  & -0.1125   & 4.1938  & 0.0048 & 0.8127  & 0.3911           & 0.8371       & 2.2030                & 3.4641                 \\ \hline
        \end{tabular}}

     \label{tb3}
\end{table}
Following \cite{hou2021extreme}, we use “loss” to represent the percentage of the negative weekly returns. Upon calculation, we have $n=1565$ observations of losses for HSI and the 17 firms. Table \ref{tb3} shows the tickers, full names and the summary statistics of the losses for the 17 firms plus HSI. Figure \ref{fg4} shows the boxplots of all losses. 

Before studying the effect of the extreme loss of HSI, the systemic variable $Y$, on the losses of individual stocks, denoted by $X_i, i=1, \cdots, 17$, we would like to check the  asymptotic independence assumption for each pair of $(X_i, Y), i=1, \cdots, 17$. Here, we apply the Tail Quotient Correlation Coefficient (TQCC)-based test in \cite{zhang2017random} to test the null hypothesis of asymptotic independence, which corresponds to the case $\eta\in(0,1)$. 

To conduct the test, we first fit generalized extreme value distribution to each series and perform marginal transformations.  We apply TQCC to the transformed data, and choose the threshold as the smaller one of the two empirical 95th percentiles. For more details on TQCC-test, we refer to \cite{zhang2017random}. The computed TQCC measures and $p$-values are summarized in Table \ref{tb3}. 
We exclude stocks with $p$-values under 0.05 from consideration. Specifically, HSBC Holdings (Ticker: X0005.HK), New World Development (Ticker: X0017.HK), and Sino Land (Ticker: X0083.HK) are removed due to substantial statistical evidence supporting the rejection of asymptotic independence between these stocks and HSI.

\begin{figure}

\begin{subfigure}[b]{0.24\textwidth}
    \centering
    \includegraphics[width=\linewidth]{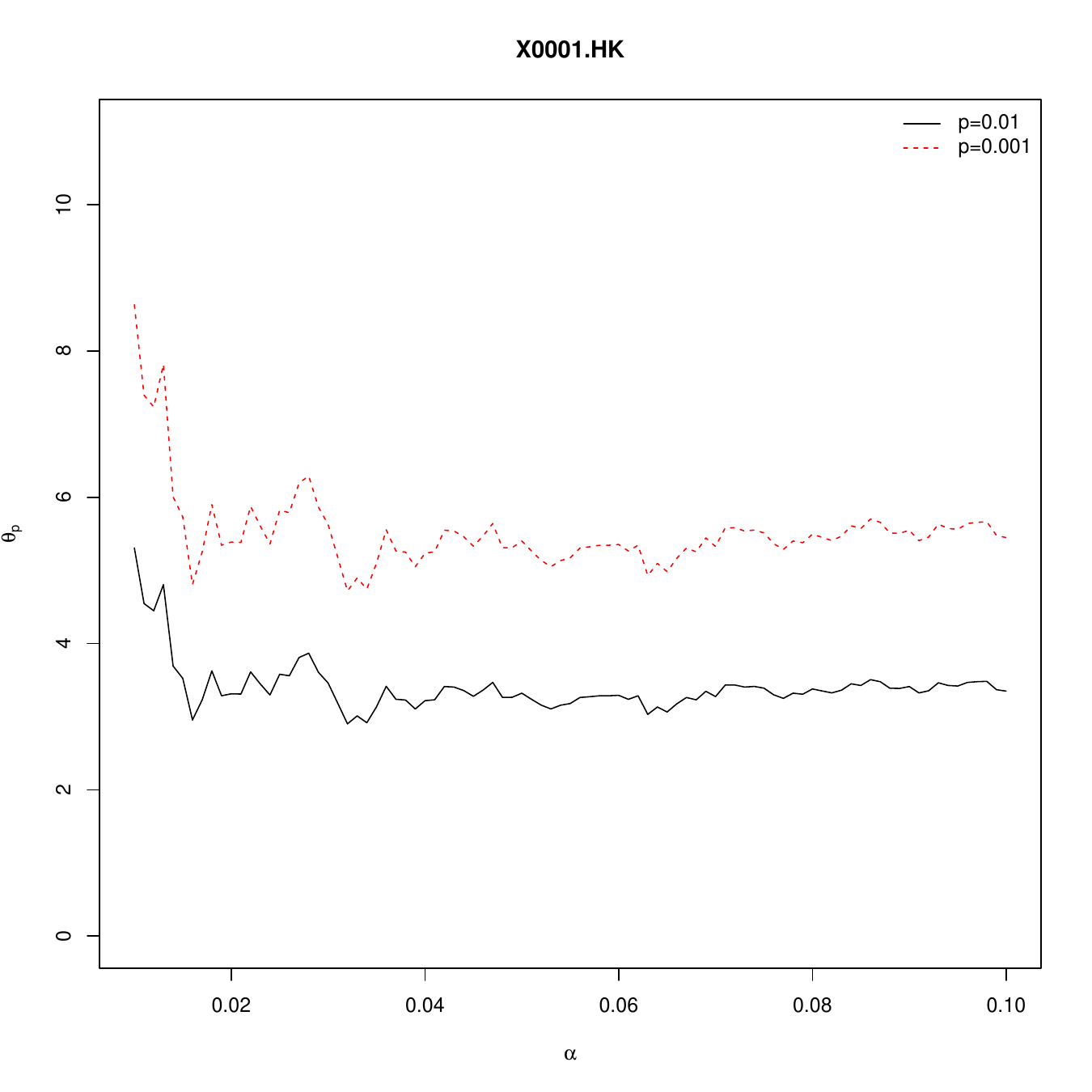}
\end{subfigure}
\hfill
\begin{subfigure}[b]{0.24\textwidth}
        \centering
        \includegraphics[width=\linewidth]{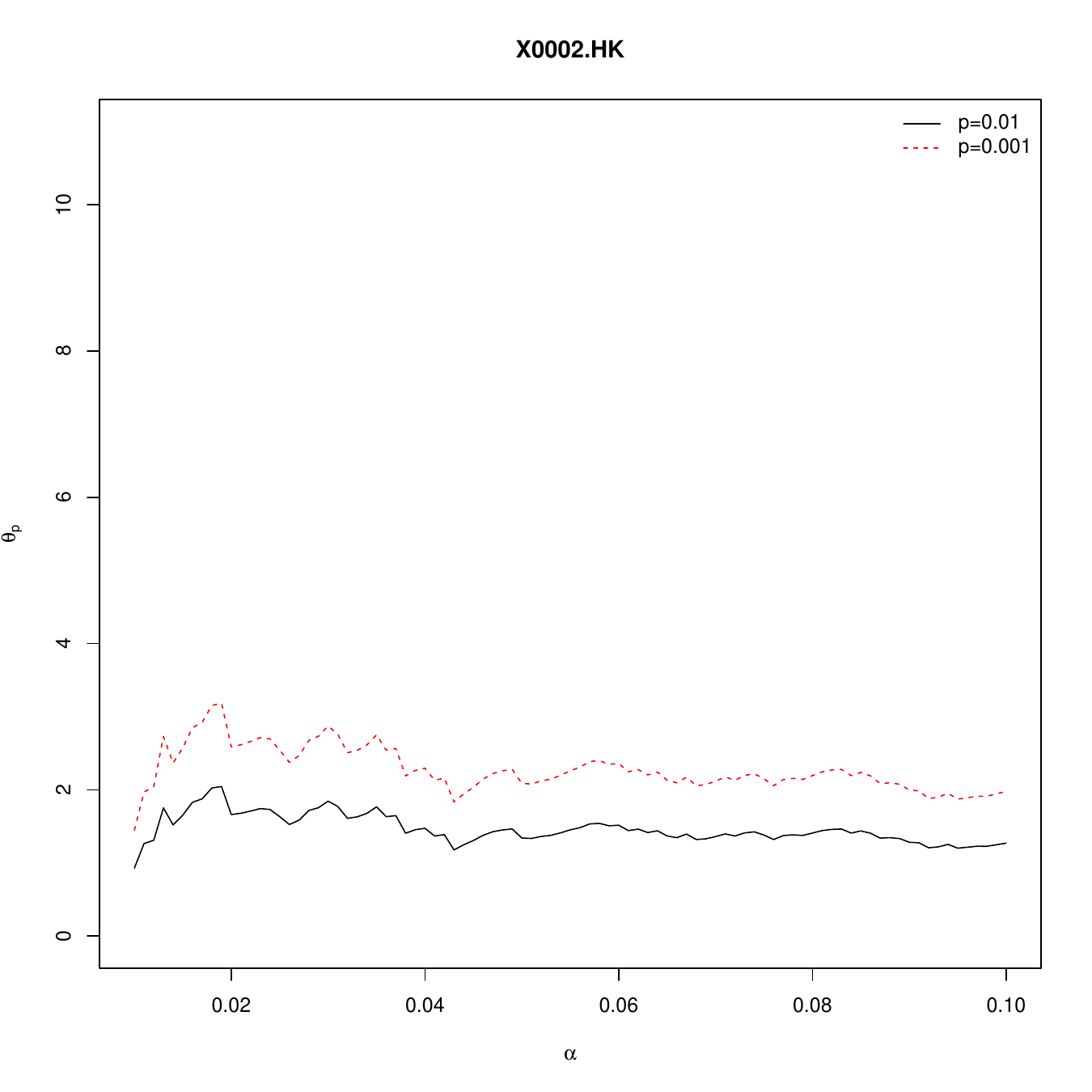}
    \end{subfigure}\hfill
    \begin{subfigure}[b]{0.24\textwidth}
        \centering
        \includegraphics[width=\linewidth]{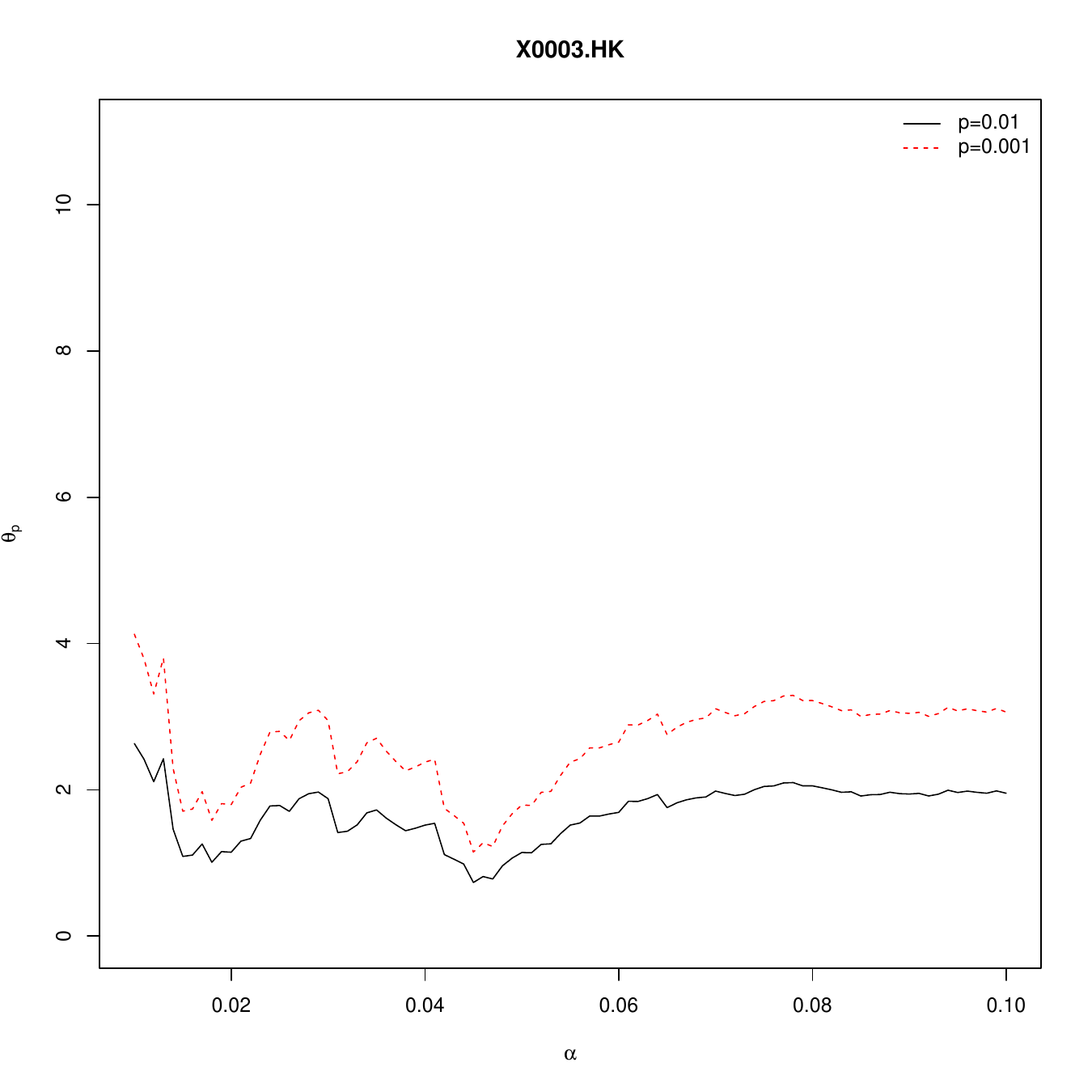}
    \end{subfigure}
    \hfill
    \begin{subfigure}[b]{0.24\textwidth}
        \centering
        \includegraphics[width=\linewidth]{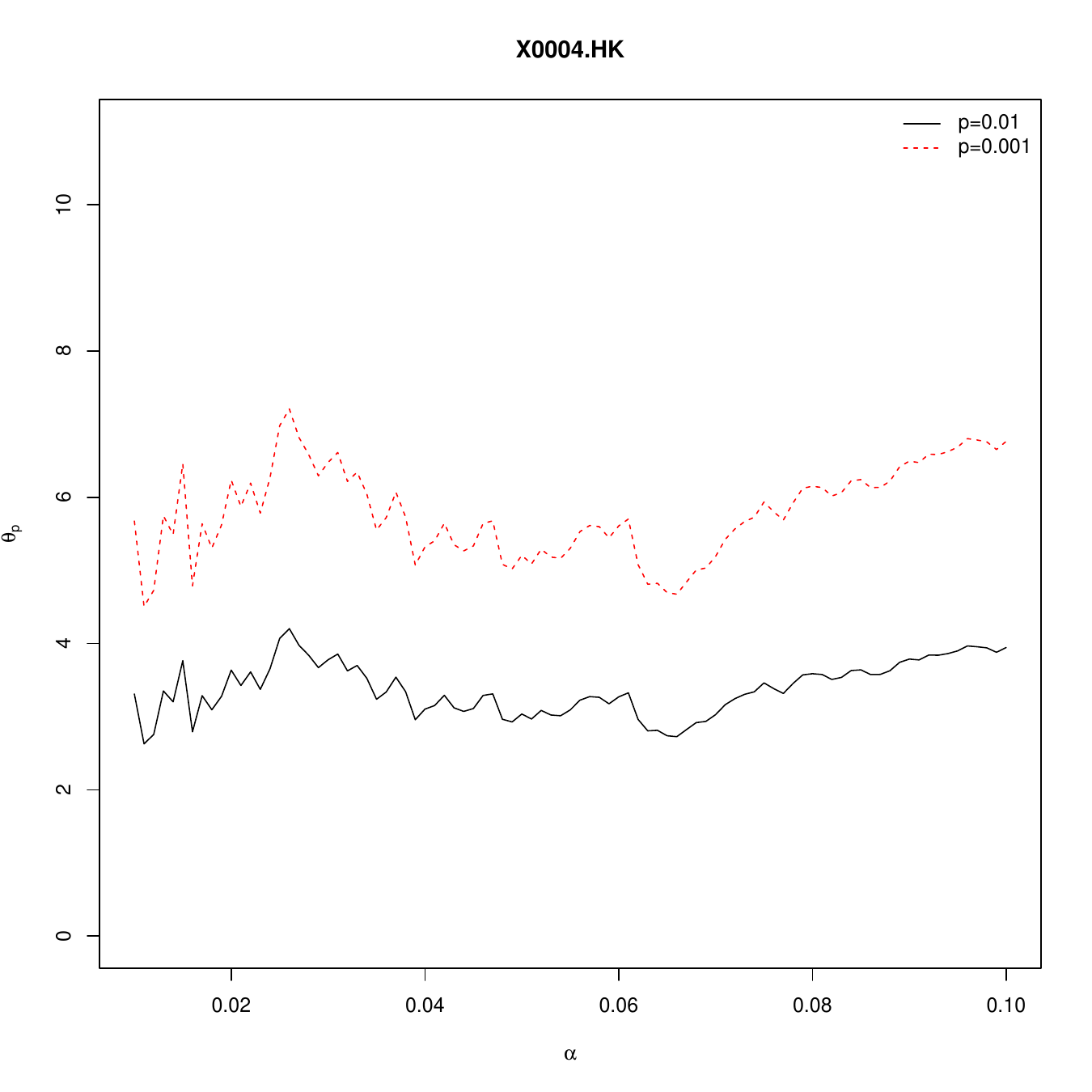}
    \end{subfigure}
    \hfill
    \begin{subfigure}[b]{0.24\textwidth}
        \centering
        \includegraphics[width=\linewidth]{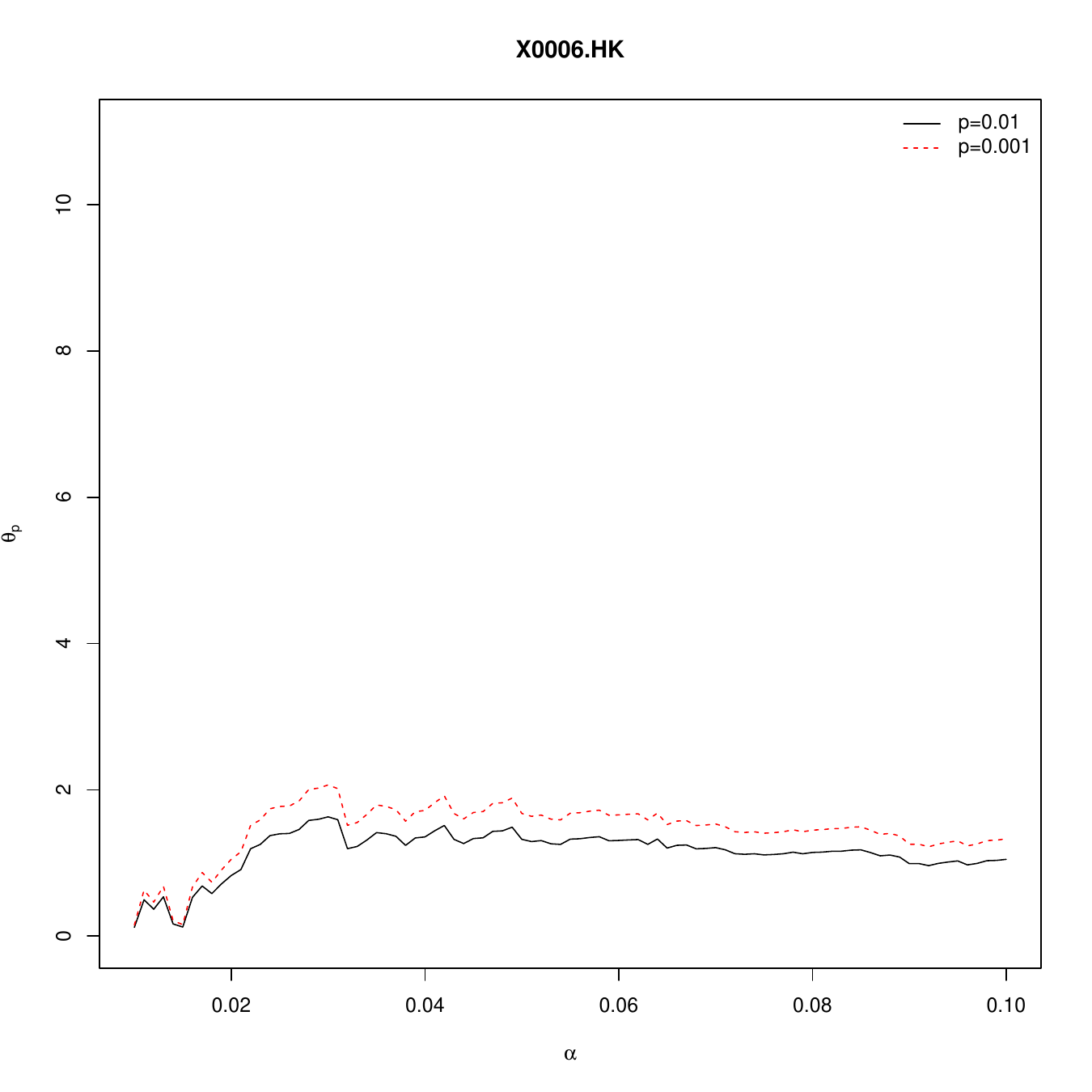}
    \end{subfigure}
    \hfill
    \begin{subfigure}[b]{0.24\textwidth}
        \centering
        \includegraphics[width=\linewidth]{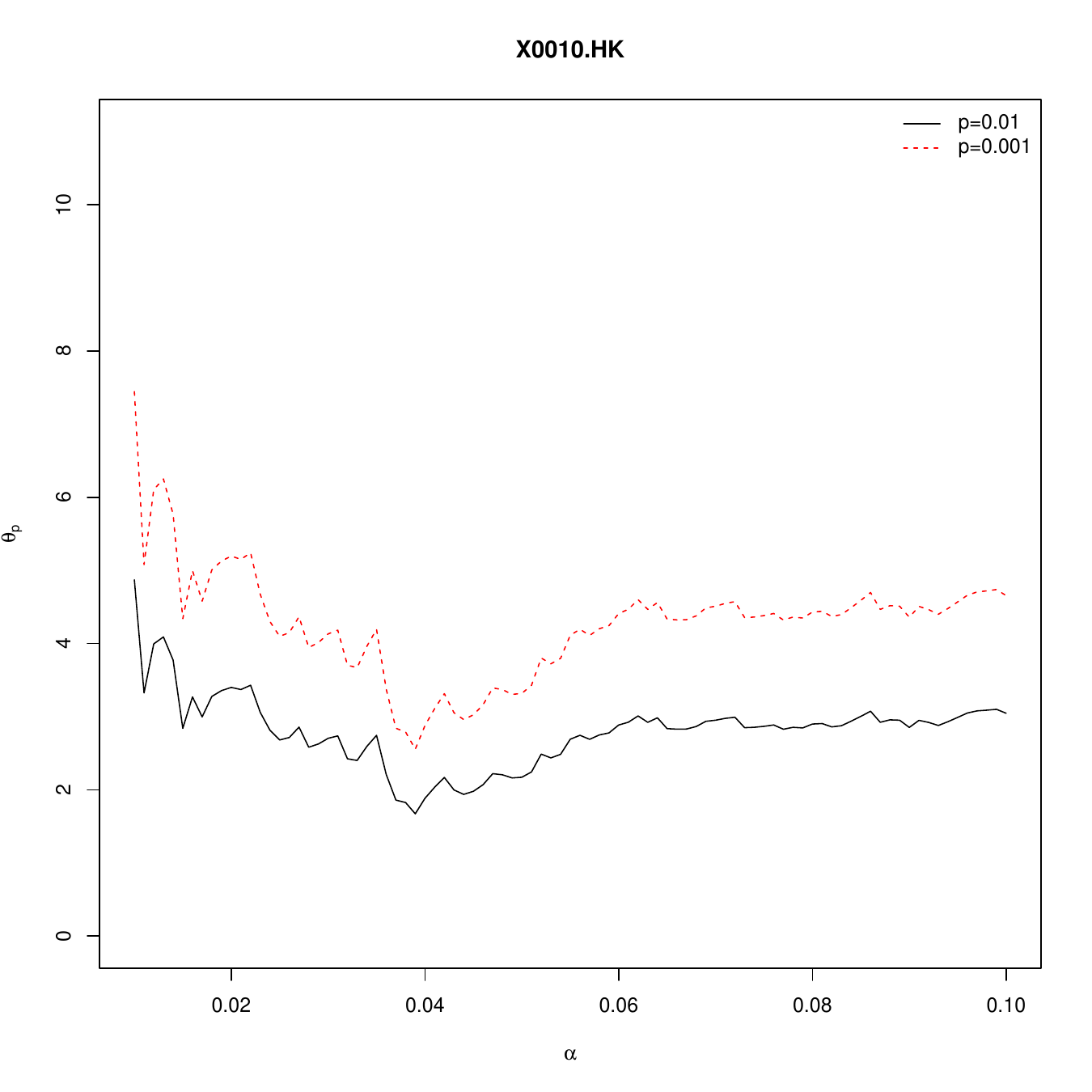}
    \end{subfigure}\hfill
    \begin{subfigure}[b]{0.24\textwidth}
        \centering
        \includegraphics[width=\linewidth]{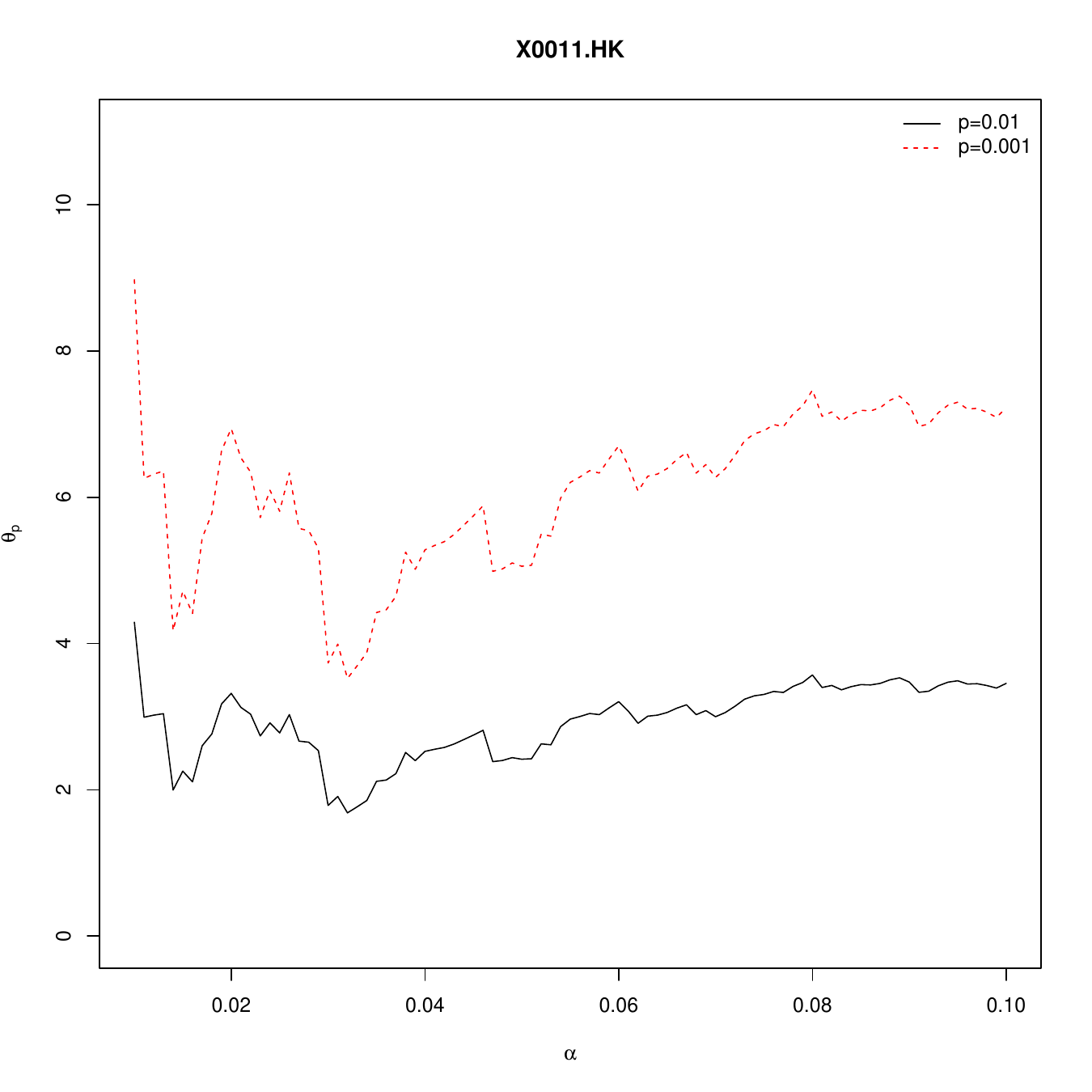}
    \end{subfigure}
    \hfill
    \begin{subfigure}[b]{0.24\textwidth}
        \centering
        \includegraphics[width=\linewidth]{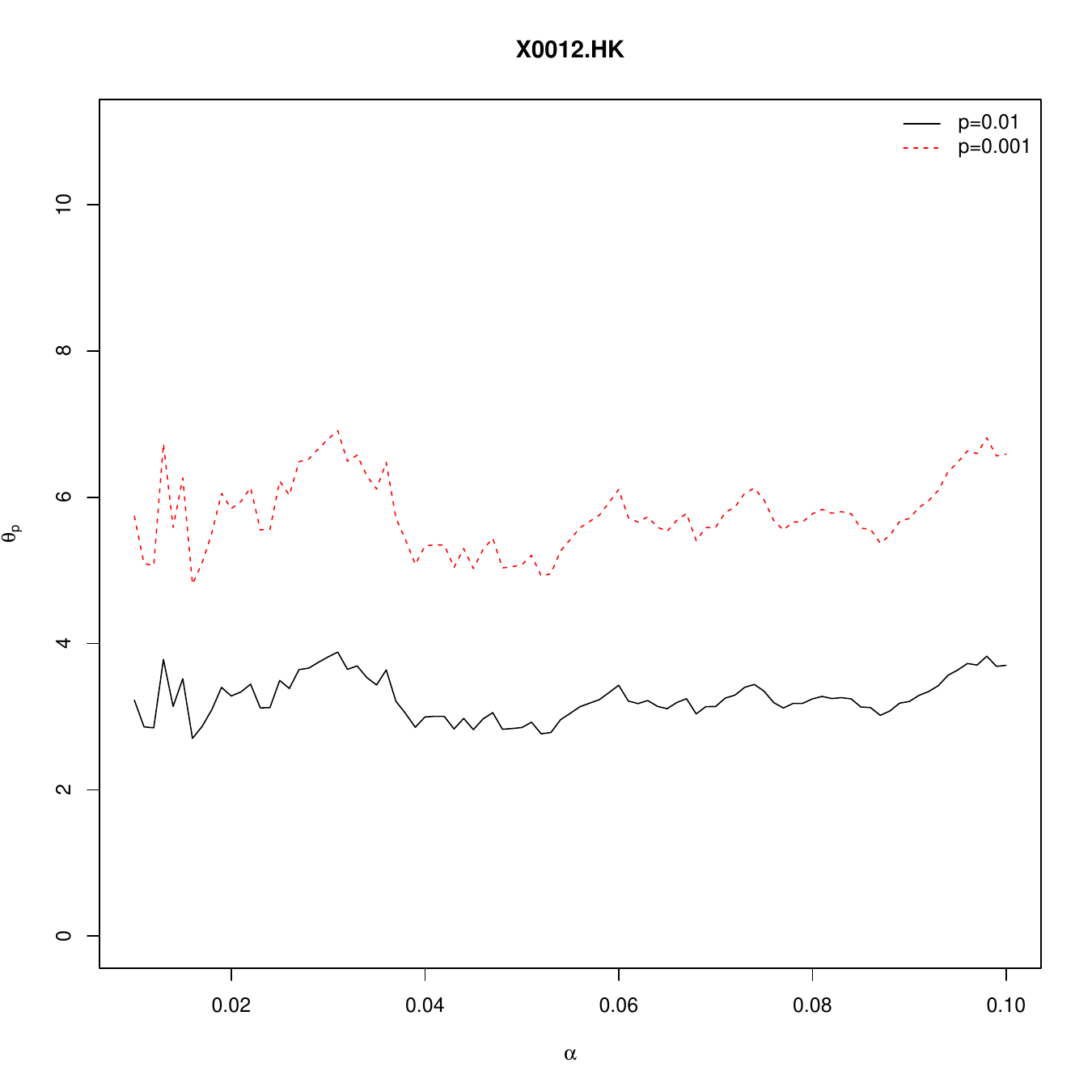}
    \end{subfigure}\hfill
    \begin{subfigure}[b]{0.24\textwidth}
        \centering
        \includegraphics[width=\linewidth]{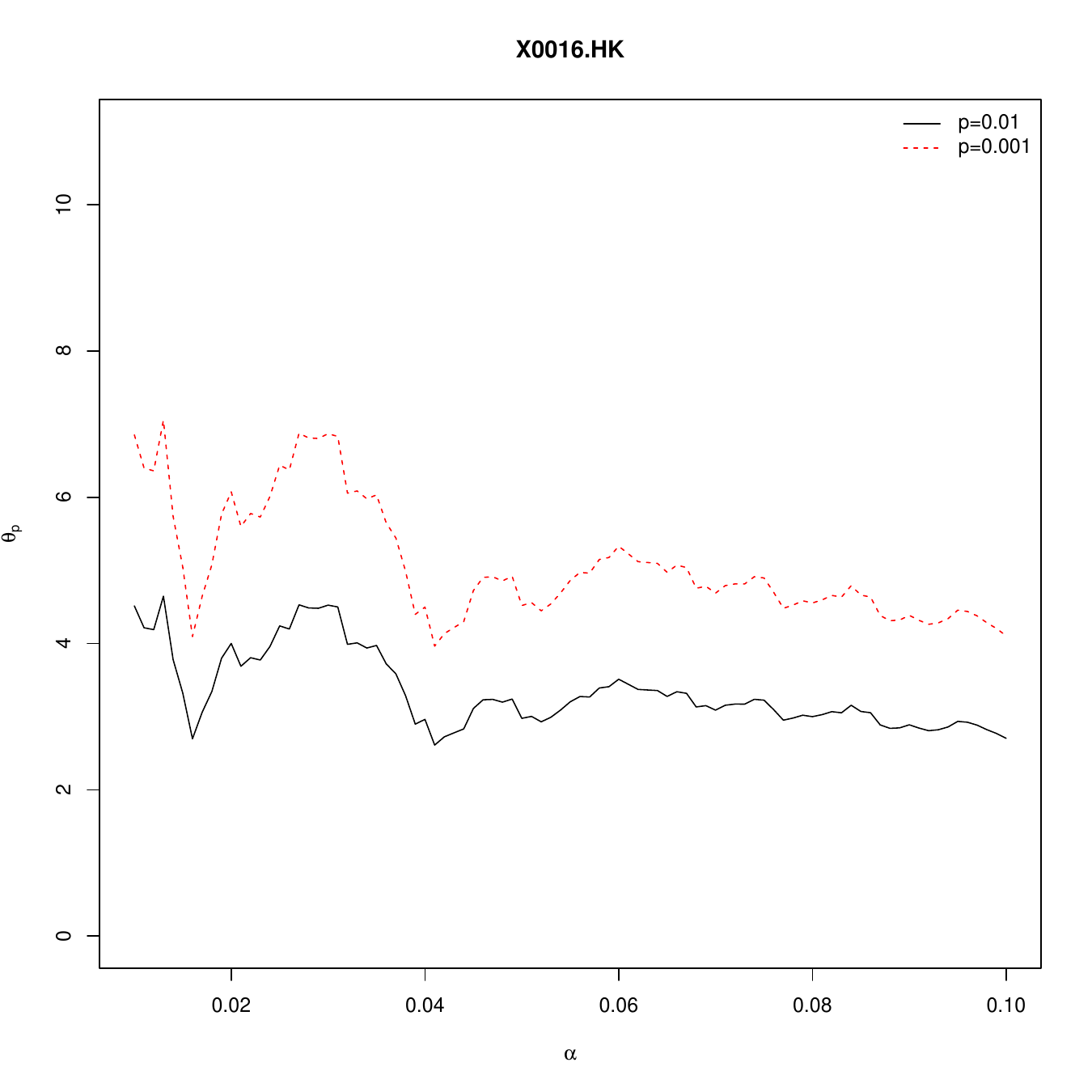}
    \end{subfigure}
    \hfill
    \begin{subfigure}[b]{0.24\textwidth}
        \centering
        \includegraphics[width=\linewidth]{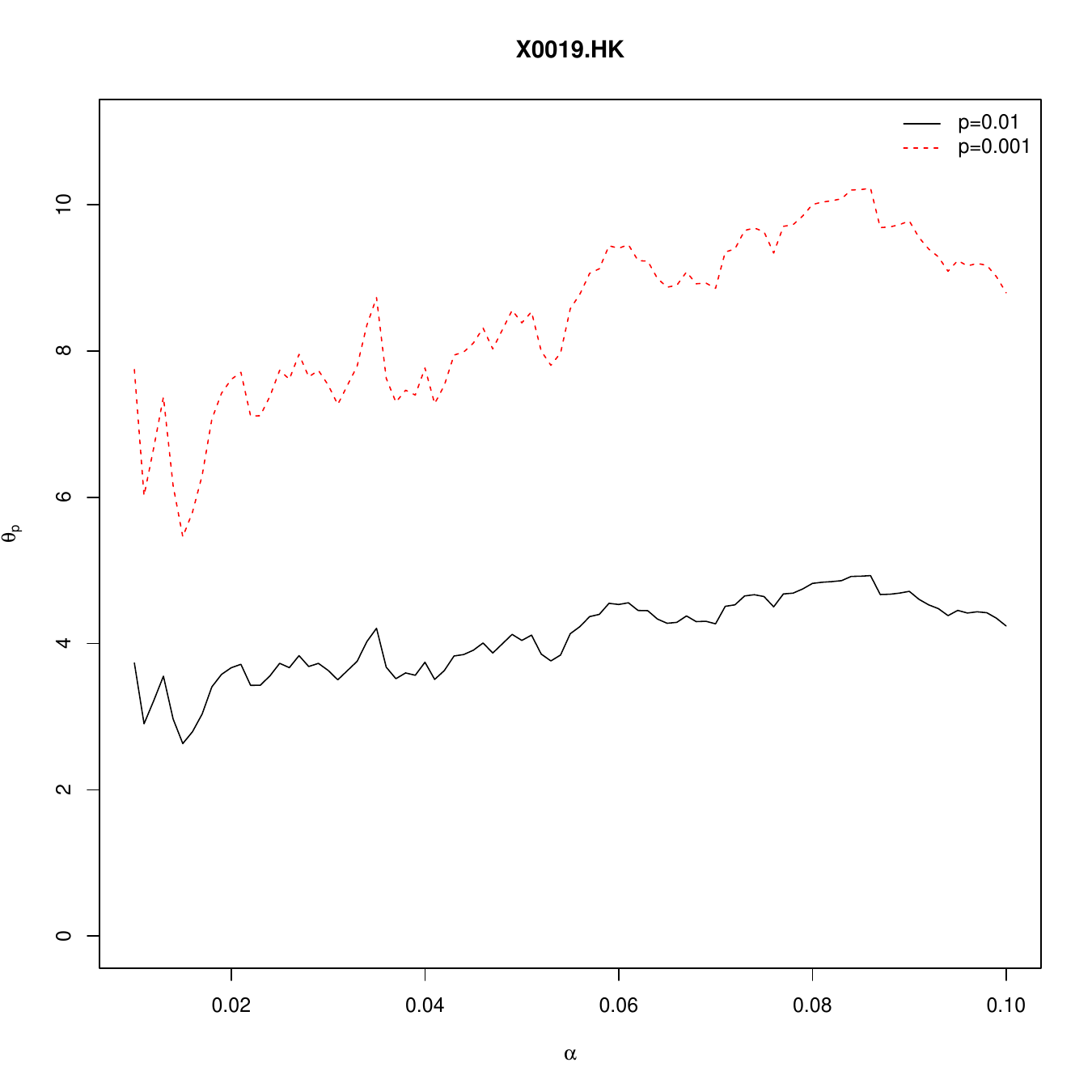}
    \end{subfigure}
    \hfill
    \begin{subfigure}[b]{0.24\textwidth}
        \centering
        \includegraphics[width=\linewidth]{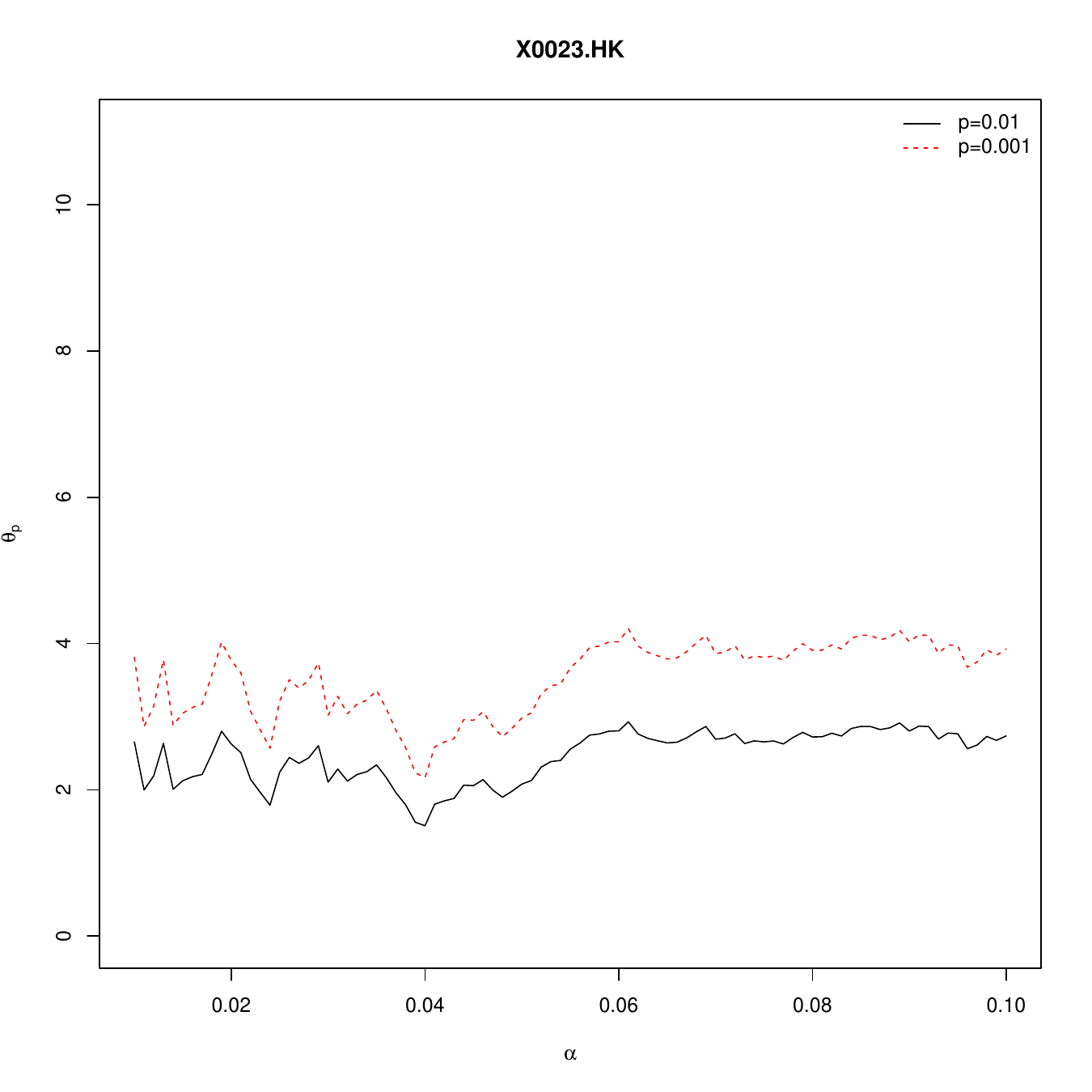}
    \end{subfigure}
    \hfill
    \begin{subfigure}[b]{0.24\textwidth}
        \centering
        \includegraphics[width=\linewidth]{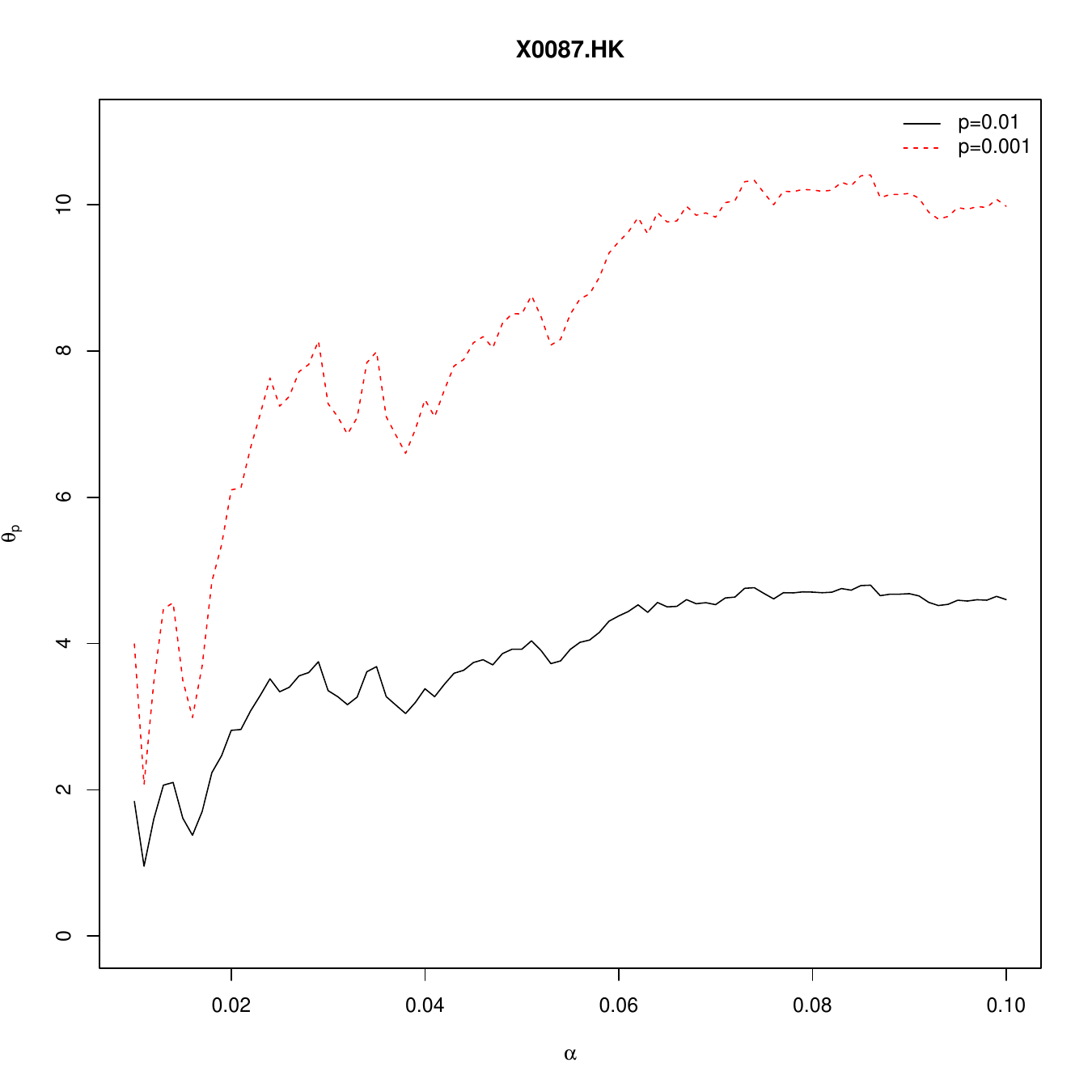}
    \end{subfigure}
    \hfill
    \begin{subfigure}[b]{0.24\textwidth}
        \centering
        \includegraphics[width=\linewidth]{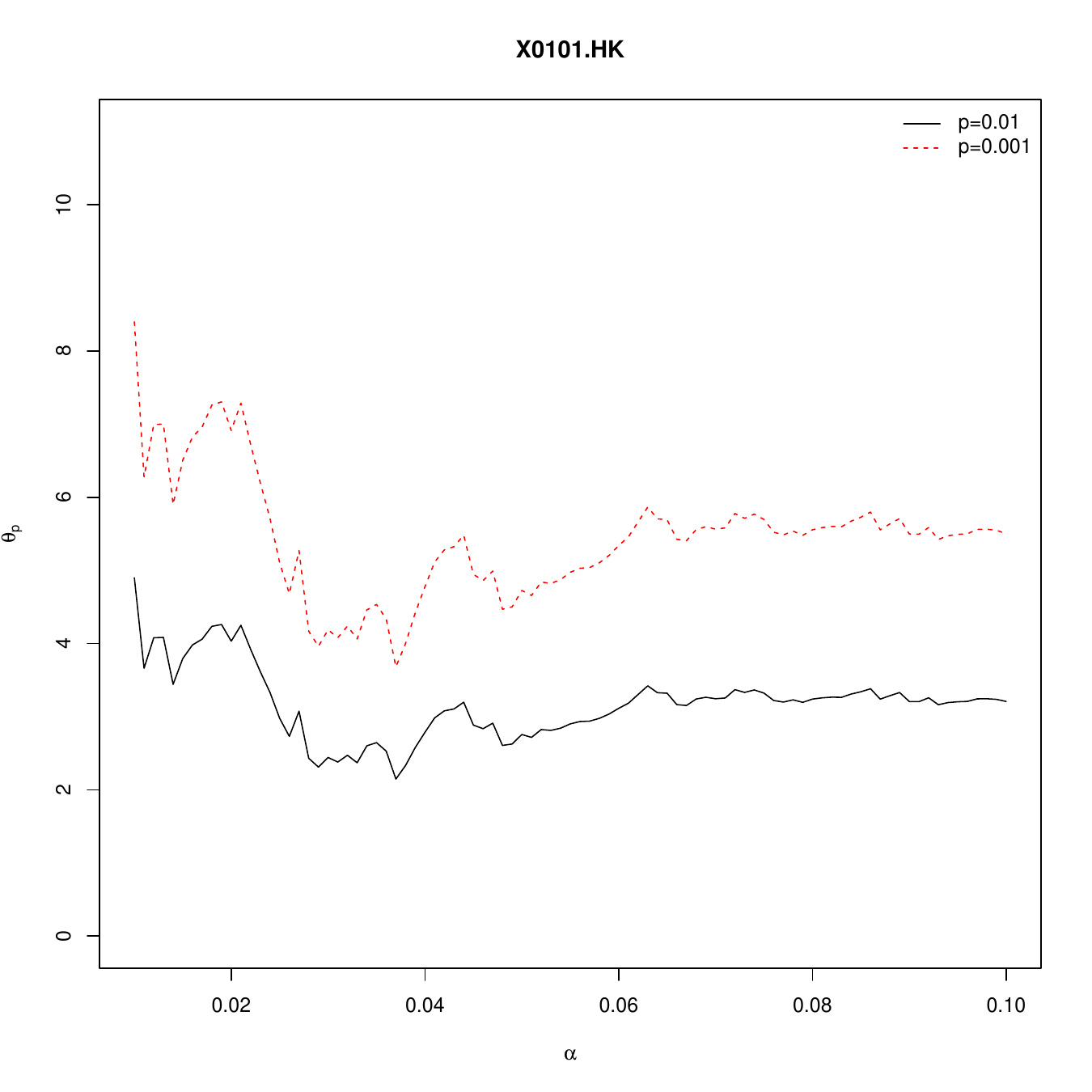}
    \end{subfigure}\hfill
    \begin{subfigure}[b]{0.24\textwidth}
        \centering
        \includegraphics[width=\linewidth]{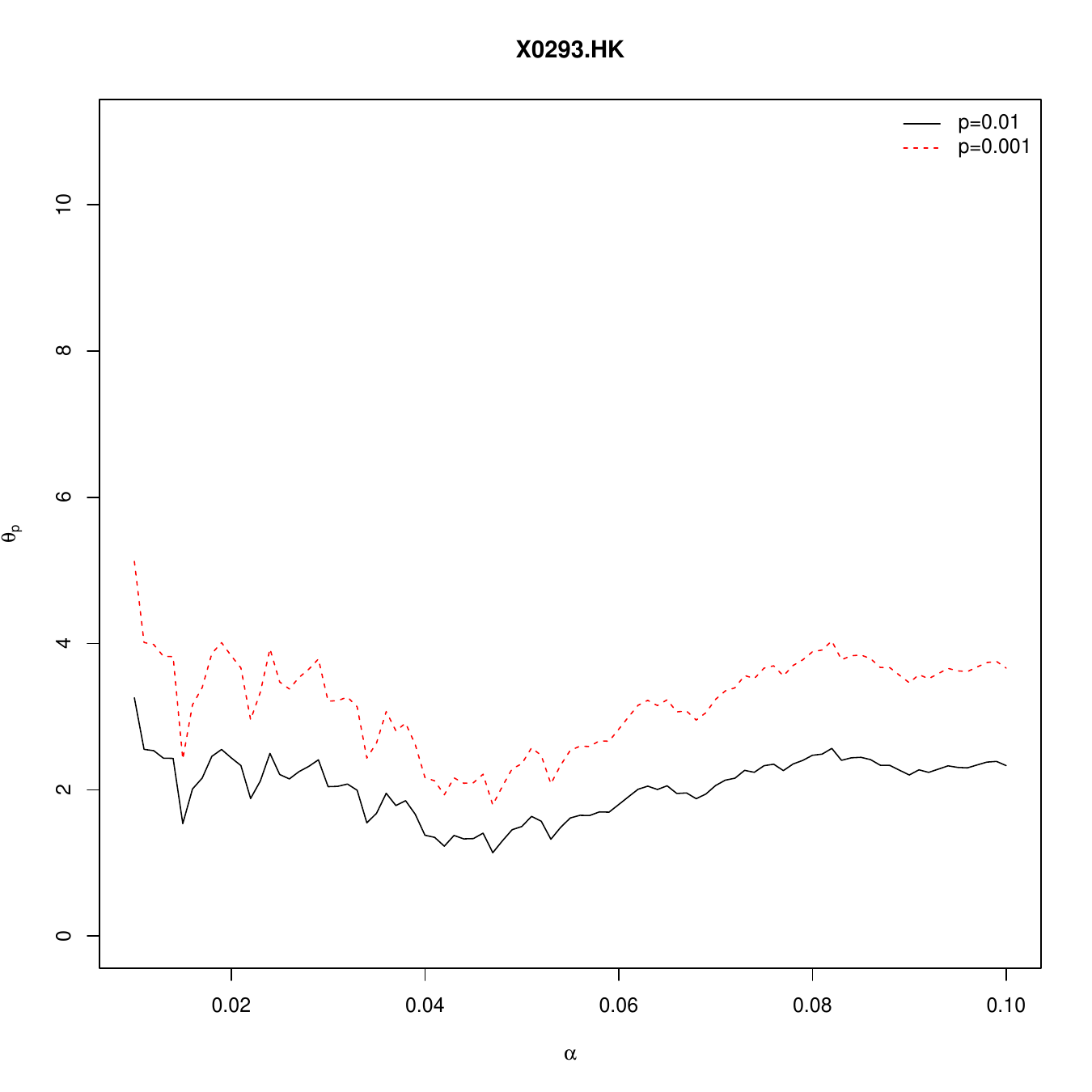}
    \end{subfigure}\hfill
     \caption{The estimates of $\operatorname{TG}_{0.01}$ and $\operatorname{TG}_{0.001}$ against $\alpha$ for the 14 stocks.}
     \label{fg7}
    \end{figure}

Subsequently, we assess the signs of $\hat{\gamma}_1$ and $\hat{\eta}$ for the remaining set of 14 stocks. From Figures S1 and S2 in the supplementary material, we can see that $\hat{\gamma}_1>0$ and $\hat{\eta}\in(0.5, 1)$ for each pair of losses across different $\alpha_1$ and $\alpha_2$. Choosing $\alpha_1 =\alpha_2 = 0.08$, we obtain the corresponding $\hat{\gamma}_1$ and $\hat{\eta}$ in Table \ref{tb3}. Figure \ref{fg7} plots the values of AIE estimator $\hat{\theta}_{0.01}$ and $\hat{\theta}_{0.001}$ against $\alpha$ for the 14 stocks, from which we conclude $\alpha = 0.09$ lying in the interval where the estimates are stable. We thus report the corresponding estimators for $p=0.01$ and $p=0.001$ in Table \ref{tb3}. It is evident that the values generated by AIE estimators exhibit a consistent pattern. They all remain below 6 when considering the scenario with $p=0.01$, and similarly, they remain below 15 when dealing with $p=0.001$. These values are notably smaller when contrasted with the findings presented in the reference \cite{hou2021extreme}. In essence, this implies that asymptotic independence typically aligns with reduced tail variability in relation to extreme events within systemic variables.

    \section{Proofs}\label{sec6}
    \textbf{Proof of Theorem \ref{thm1}.} Let $M_n=\sqrt{k}\left(\frac{n}{k}\right)^{-\frac{1}{2 \eta}+\frac{1}{2}}$. Recall that $d_n=\frac{k}{np}$, $\hat{\theta}_p=(\frac{k}{np})^{1-1/\hat{\eta}+\hat{\gamma}_1}\hat{\theta}_{k/n}$. 
    
       We rewrite\begin{align*}\frac{\hat{\theta}_p}{\mathrm{TG}_p(X ; Y)}&=d_n^{\hat{\gamma}_1-\gamma_1}  \times d_n^{\frac{1}{\eta}-\frac{1}{\hat{\eta}}} \times \frac{\hat{\theta}_{k/n}}{\mathrm{TG}_{k / n}(X ; Y)}  \times\frac{d_n^{-\frac{1}{\eta}+1} \mathrm{TG}_{k / n}(X ; Y) / Q_1(1-k / n)}{\mathrm{TG}_p(X ; Y) / Q_1(1-p)} \times\frac{d_n^{\gamma_1} Q_1(1-k / n)}{Q_1(1-p)} .
                   \\&=: I_1 \times I_2 \times I_3 \times I_4\times I_5 .
               \end{align*}
               By the assumption that $M_n\log d_n=o(\sqrt{k})$ and $\sqrt{k}\left(\hat{\gamma}_1-\gamma_1\right)=O_{\mathbf{P}}(1)$, it follows that
               $$
               I_1-1=e^{\left(\hat{\gamma}_1-\gamma_1\right) \log d_n}-1=\left(\hat{\gamma}_1-\gamma_1\right) \log d_n+o_{\mathbf{P}}\left(\left(\hat{\gamma}_1-\gamma_1\right) \log d_n\right)=O_{\mathbf{P}}\left(\frac{\log d_n}{\sqrt{k}}\right)=o_{\mathbf{P}}\left(\frac{1}{M_n}\right).
               $$In the same way, we get $I_2-1=o_{\mathbf{P}}\left(\frac{1}{M_n}\right)$. By Proposition \ref{prop2}  we have
               $$
               I_3=\frac{\hat{\theta}_{k/n}}{\mathrm{TG}_{k / n}(X ; Y)}=1+\frac{1}{M_n} \Phi+o_{\mathbf{P}}\left(\frac{1}{M_n}\right) .
               $$

               For $I_4$, by (4), (ii) and (iv) in Lemma 1 in supplementary material, we have \begin{align*}\frac{\operatorname{TG}_{k / n}(X; Y)}{(\frac{k}{n})^{\frac{1}{\eta}-1} Q_1(1-k/n)}=&4\int_0^{\infty} \int_0^1\tau_{k/n}\left(s_{k/n}(x),y\right)d y d x^{-\gamma_1}-2\int_0^{\infty} \tau_{k/n}\left(s_{k/n}(x),1\right)d x^{-\gamma_1}\\=&\phi_0+o\left(\frac{1}{M_n}\right).\end{align*}Similarly, for $p\leq k/n$, we have $$\frac{\operatorname{TG}_{p}(X; Y)}{p^{\frac{1}{\eta}-1} Q_1(1-p)}=\phi_0+o\left(\frac{1}{M_n}\right).$$Thus 
               $$I_4=\frac{(k/n)^{1-\frac{1}{\eta}} \operatorname{TG}_{k / n}(X ; Y) / Q_1(1-k / n)}{p^{1-\frac{1}{\eta}}\operatorname{TG}_p(X ; Y) / Q_1(1-p)}=\frac{\phi_0+o\left(\frac{1}{M_n}\right)}{\phi_0+o\left(\frac{1}{M_n}\right)}=1+o\left(\frac{1}{M_n}\right).
               $$It follows from Assumptions \ref{asm1} and \ref{asm5} that
               $$
               I_5=\frac{d_n^{\gamma_1} Q_1(1-k / n)}{Q_1(1-p)}=1+O\left(A_1\left(\frac{n}{k}\right)\right) =1+o(\frac{1}{\sqrt{k}}).
               $$
               
               Hence, we finally obtain\begin{align*}\frac{\hat{\theta}_p}{\mathrm{TG}_p(X ; Y)}&=\left(1+o_{\mathbf{P}}\left(\frac{1}{M_n}\right)\right)^2\times \left(1+\frac{\Phi}{M_n} +o_{\mathbf{P}}\left(\frac{1}{M_n}\right)\right)\times \left(1+o\left(\frac{1}{M_n}\right)\right)\times \left(1+o(\frac{1}{\sqrt{k}})\right)
                   \\&=1+\frac{\Phi}{M_n}+o_{\mathbf{P}}\left(\frac{1}{M_n}\right),
               \end{align*}
       which implies the statement of Theorem \ref{thm1}.$\hfill\qedsymbol$
       \bigskip \\\textbf{Proof of Theorem \ref{thm2}.} Write $$\frac{\hat{\theta}_{p}}{\mathrm{TG}_{p}(X ; Y)}=\frac{\hat{\theta}_p}{\mathrm{TG}_{p}(X^{+} ; Y)}\times \frac{\mathrm{TG}_{p}(X^{+} ; Y)}{\mathrm{TG}_{p}(X ; Y)}.$$We first consider $\frac{\hat{\theta}_p}{\mathrm{TG}_{p}(X^{+} ; Y)}$ and show that it has the same the asymptotic normality as $\frac{\hat{\theta}_p}{\mathrm{TG}_{p}(X; Y)}$, which is stated in Theorem \ref{thm1}. In other words, we need to check that Assumptions \ref{asm1} to \ref{asm6} also hold for $\left(X^{+}, Y\right)$. Note that Assumptions \ref{asm2}, \ref{asm5} and \ref{asm6} hold automatically. Thus we only need to show that $\left(X^{+}, Y\right)$ satisfies Assumptions \ref{asm1}, \ref{asm3} and \ref{asm4}. 
       
       Denote the distribution of $X^{+}$as $F_1^{+}$, the quantile function of $X^{+}$as $Q_1^{+}$,  and $$\tau^+_p(x, y)=p^{-1/\eta} \mathbf{P}\left(1-F_1^+(X^+)<px, 1-F_2(Y)<py\right), \quad x, y>0.$$ As $X$ has a continuous distribution, a simple calculation leads to ${Q}_1^{+}(1-p)=Q_1(1-p) I(0<p\leq \bar{F}_1(0))$, which implies that $X^+$ satisfies Assumption \ref{asm1} and that
       $$\mathbf{P}\left(1-F_1^{+}\left(X^{+}\right)<u, 1-F_2(Y)<v\right)= \left\{ \begin{array}{ll}\mathbf{P}\left(1-F_1\left(X\right)<u, 1-F_2(Y)<v\right), & 0<u<\bar{F}_1(0), \\ v,& u\geq\bar{F}_1(0). \end{array} \right.$$Thus, $$
       \tau_p^{+}(x, y)= \left\{ \begin{array}{ll}
        \tau_p(x, y), & 0<x<\bar{F}_1(0) / p,\\
        p^{1-\frac{1}{\eta}} y, & x \geq \bar{F}_1(0) / p,
         \end{array} \right.
       $$which means that Assumption \ref{asm3} also holds for $\left(X^{+}, Y\right)$. 
       
       For Assumption \ref{asm4}, notice that for suffiently small $p>0$, $$\sup _{\substack{1<x<\bar{F}_1(0)/p  \\ 0 \leq y \leq 1}}\left|\tau_p^+(x, y)-\tau(x, y)\right| x^{-\beta_2}=\sup _{\substack{1<x<\bar{F}_1(0)/p \\ 0 \leq y \leq 1}}\left|\tau_p(x, y)-\tau(x, y)\right| x^{-\beta_2}=O\left(p^{\xi}\right).$$Moreover, by the homogeneity of $\tau$ and setting $x =1/p$, we have, for $0<y\leq 1$\begin{equation}p^{1-1/\eta}y-y^{1/\eta}\tau(\frac{1}{py},1)=O(p^{\xi-\beta_2}).\label{6}\end{equation}So for $0<y\leq 1$, we have $$p^{1-\frac{1}{\eta}}y-y^{1/\eta}\tau(\frac{\bar{F}_1(0)}{py}, 1)=\left\{ \begin{array}{ll}O(p^{1-1/\eta}), & 1-1/\eta<\xi-\beta_2, \\ O(p^{\xi-\beta_2}), & 1-1/\eta>\xi-\beta_2. \end{array} \right.$$Therefore, for $x\geq\bar{F}_1(0)/p$ and $p$ suffiently small, it follows that\begin{align*}x^{-\beta_2}\left(\tau^+_p(x, y)-\tau(x, y)\right)&=x^{-\beta_2}\left(p^{1-\frac{1}{\eta}}y-\tau(x, y)\right)\\&\leq \left(\frac{\bar{F}_1(0)}{p}\right)^{-\beta_2}\left(p^{1-\frac{1}{\eta}}y-\tau(\bar{F}_1(0)/p, y)\right)\\&= \left(\frac{p}{\bar{F}_1(0)}\right)^{\beta_2}\left(p^{1-\frac{1}{\eta}}y-y^{1/\eta}\tau(\frac{\bar{F}_1(0)}{py}, 1)\right)\\&=\left\{ \begin{array}{ll}O(p^{\beta_2+1-1/\eta})
           , & 1-1/\eta<\xi-\beta_2, \\ O(p^{\xi}), & 1-1/\eta>\xi-\beta_2. \end{array} \right.
       \end{align*}Since $1-1/\eta>\xi-\beta_2$ (see Assumption \ref{asm7}), it follows that$$\sup _{\substack{x\geq\bar{F}_1(0)/p  \\ 0 \leq y \leq 1}}\left|\tau_p^+(x, y)-\tau(x, y)\right| x^{-\beta_2}=O\left(p^{\xi}\right),$$which means that Assumption \ref{asm4} also holds for $(X^{+} ; Y)$.
       
       As a result, Theorem \ref{thm1} can be applied and we have$$
       \sqrt{k}\left(\frac{n}{k}\right)^{-\frac{1}{2 \eta}+\frac{1}{2}}\left(\frac{\hat{\theta}_p}{\operatorname{TG}_p(X^{+}; Y)}-1\right) \stackrel{d}{\rightarrow} \Phi.
       $$ 
       
       Next we show $\frac{\mathrm{TG}_{p}(X ; Y)}{\mathrm{TG}_{p}(X^{+} ; Y)}=1+o\left(\frac{1}{M_n}\right)$. Note that $$\frac{\mathrm{TG}_{p}(X ; Y)}{\mathrm{TG}_{p}(X^{+} ; Y)}-1=\frac{4}{p}\frac{\operatorname{Cov}\left(X^{-}, F_2(Y) \mid F_2(Y)>1-p\right)}{\mathrm{TG}_{p}(X^{+} ; Y)}.$$Rewrite
       \begin{align*}&\operatorname{Cov}\left(X^{-}, F_2(Y) \mid F_2(Y)>1-p\right)\\=&\mathbb{E}\left(X^{-}F_2(Y) \mid F_2(Y)>1-p\right)-\mathbb{E}\left(X^{-} \mid F_2(Y)>1-p\right)\mathbb{E}\left(F_2(Y) \mid F_2(Y)>1-p\right)\\\leq &\mathbb{E}\left(|X^{-}|F_2(Y) \mid F_2(Y)>1-p\right)+\mathbb{E}\left(|X^{-}| \mid F_2(Y)>1-p\right)\mathbb{E}\left(F_2(Y) \mid F_2(Y)>1-p\right)\\\leq &2\mathbb{E}\left(|X^{-}| \mid F_2(Y)>1-p\right)\\=&\frac{2}{p}\mathbb{E}\left(|X^{-}|I\left(X<0, \bar{F}_2(Y)<p\right)\right)\\\leq&\frac{2}{p}\left(\mathbb{E}\left|X^{-}\right|^{\zeta}\right)^{1/\zeta}\left(\mathbf{P}\left(X<0, \bar{F}_2(Y)<p\right)\right)^{1-1/\zeta},\end{align*}where the last inequality is guaranteed by  H\"older's inequality. By Assumption \ref{asm7}, we have $\mathbb{E}\left|X^{-}\right|^{\zeta}<\infty$. 
       
       Now we handle $\mathbf{P}\left(X<0, \bar{F}_2(Y)<p\right)$, as $p\to0$. Note that \begin{align*}\mathbf{P}\left(X<0, \bar{F}_2(Y)<p\right)=& p^{1/\eta}\left(p^{1-1/\eta}-\tau_p\left(\bar{F}_1(0)/p, 1\right)\right) \\=&p^{1/\eta}\left(p^{1-1/\eta}-\tau\left(\bar{F}_1(0)/p, 1\right)+\tau\left(\bar{F}_1(0)/p, 1\right)-\tau_p\left(\bar{F}_1(0)/p, 1\right)\right). \end{align*}
       By Assumption \ref{asm4},$$\tau\left(\bar{F}_1(0)/p, 1\right)-\tau_p\left(\bar{F}_1(0)/p, 1\right)=O(p^{\xi-\beta_2}).$$ Setting $y=1/\bar{F}_1(0)$ in (\ref{6}), we have $$p^{1-1/\eta}/\bar{F}_1(0)-\bar{F}_1(0)^{-1/\eta}\tau(\bar{F}_1(0)/p,1)=O(p^{\xi-\beta_2}).$$Recall that $1-1/\eta>\xi-\beta_2$. Then, \begin{align*}p^{1-1/\eta}-\tau\left(\bar{F}_1(0)/p, 1\right)=&\bar{F}_1(0)^{1/\eta}\Big(p^{1-1/\eta}/\bar{F}_1(0)-\bar{F}_1(0)^{-1/\eta}\tau(\bar{F}_1(0)/p,1)\Big)+\Big(1-\bar{F}_1(0)^{1/\eta-1}\Big)p^{1-1/\eta}\\=&O(p^{\xi-\beta_2}),
       \end{align*}
           which leads to$$\mathbf{P}\left(X<0, \bar{F}_2(Y)<p\right)=
           O(p^{1/\eta+\xi-\beta_2}),
           $$and$$\operatorname{Cov}\left(X^{-}, F_2(Y) \mid F_2(Y)>1-p\right)=
               O(p^{(1/\eta+\xi-\beta_2)(1-1/\zeta)-1}). 
               $$
               Thus, it follows that\begin{align*}M_n\left(\frac{\mathrm{TG}_{p}(X ; Y)}{\mathrm{TG}_{p}(X^{+} ; Y)}-1\right)&=M_n\frac{4}{p}\frac{\operatorname{Cov}\left(X^{-}, F_2(Y) \mid F_2(Y)>1-p\right)}{\mathrm{TG}_{p}^{+}(X ; Y)}\\&=M_n\frac{4}{p}\frac{\operatorname{Cov}\left(X^{-}, F_2(Y) \mid F_2(Y)>1-p\right)}{O(p^{1/\eta-1} Q_1(1-p))}\\&=(O(M_np^{(1-\frac{1}{\zeta})(\frac{1}{\eta}-\beta_2+\xi)-1-\frac{1}{\eta}+\gamma_1}))\\&=o(1),\end{align*}
               and the proof is completed.
       
       $\hfill\qedsymbol$
\bibliographystyle{apalike}

\bibliography{ref}
\end{document}